\documentclass[%
 preprint,
 amsmath,amssymb,
 aps,
 pra,
 12pt
]{revtex4-1}

\usepackage{graphicx}
\usepackage{subfigure}
\usepackage{bm}
\usepackage[mathlines]{lineno}
\usepackage{color}
\usepackage{psfrag}
\usepackage{hyperref}
\usepackage {CJK}
\usepackage{latexsym}
\usepackage{amssymb}
\usepackage{indentfirst}
\usepackage[letterpaper]{geometry}
\usepackage{euscript}
\usepackage{fullpage}
\usepackage{mathrsfs}
\usepackage{amsfonts}

\usepackage{braket}

\usepackage{verbatim}
\usepackage{makecell}
\usepackage[ruled]{algorithm2e}
\usepackage{amsthm}
\usepackage{multirow}
\definecolor{darkgreen}{rgb}{0.0, 0.5, 0.0}

\graphicspath{{figure/}}

\newcommand{\EQ}{\begin{equation}}
\newcommand{\EN}{\end{equation}}
\newcommand{\EQA}{\begin{eqnarray}}
\newcommand{\ENA}{\end{eqnarray}}

\newcommand{\figref}[1]{Fig.~\ref{#1}}
\renewcommand{\eqref}[1]{Eq.~(\ref{#1})}

\begin{document}

\begin{CJK*}{UTF8}{gbsn}

\title{Efficient quantum state tomography with Chebyshev polynomials}
\author{Hao Su} 
 \affiliation{State Key Laboratory for Turbulence and Complex Systems, School of Mechanics and Engineering Science, Peking University, Beijing 100871, China}
\author{Shiying Xiong} 
\email{shiying.xiong@zju.edu.cn}
 \affiliation{Department of Engineering Mechanics, School of Aeronautics and Astronautics, Zhejiang University, Hangzhou 310027, China}
\author{Yue Yang} 
 \email{yyg@pku.edu.cn}
 \affiliation{State Key Laboratory for Turbulence and Complex Systems, School of Mechanics and Engineering Science, Peking University, Beijing 100871, China}%
 \affiliation{HEDPS-CAPT, Peking University, Beijing 100871, China}

\date{\today}

\begin{abstract}
Quantum computing shows promise for addressing computationally intensive problems but is constrained by the exponential resource requirements of general quantum state tomography (QST), which fully characterizes quantum states through parameter estimation. We introduce the QST with Chebyshev polynomials, an approximate tomography method for pure quantum states encoding complex-valued functions. This method reformulates tomography as the estimation of Chebyshev expansion coefficients, expressed as inner products between the target quantum state and Chebyshev basis functions, measured using the Hadamard test circuit. By treating the truncation order of the Chebyshev polynomials as a controllable parameter, the method provides a practical balance between efficiency and accuracy. For quantum states encoding functions dominated by large-scale features, such as those representing fluid flow fields, appropriate truncation enables faithful reconstruction of the dominant components via quantum circuits with linear depth, while keeping both measurement repetitions and post-processing independent of qubit count, in contrast to the exponential scaling of full measurement-based QST methods. Validation on analytic functions and numerically generated flow-field data demonstrates accurate reconstruction and effective extraction of large-scale features, indicating the method's suitability for systems governed by macroscopic dynamics.
\end{abstract}

\maketitle
\end{CJK*}


\section{Introduction}
Quantum computing has drawn increasing attention in recent years, owing to its capacity for exponential data storage and its potential to accelerate various algorithms ~\cite{Nielsen2010, Wittek2014, Cao2019, Jin2023, Meng2024, Xiao2024}. For classical physical systems, quantum computing offers a pathway to enhance the numerical solution of models such as differential equations by mapping their discretized forms into quantum-linear algebraic problems, where quantum algorithms may provide advantages in high-dimensional scaling, sampling efficiency, and measurement strategies tailored to specific objectives. In particular, quantum computing for fluid dynamics (QCFD) is an emerging field that seeks to solve classical fluid mechanics problems using quantum approaches~\cite{Succi2023, Bharadwaj2025, Tennie2025, Meng2025}. Despite the development of various frameworks and algorithms~\cite{Steijl2018, Gaitan2020, Lubasch2020, Gourianov2022, Pfeffer2022, Zylberman2022, Fukagata2022, Meng2023, Meng2024quantum, Jaksch2023, Liu2024variational, Chen2024, Bharadwaj2023, Itani2024, Wang2025}, few can be effectively implemented on current quantum devices, partly restrained by the quantum state tomography (QST) complexity.

A quantum state fully characterizes a system, with pure states described as vectors in Hilbert space and general states represented by density matrices that encode both pure and mixed states. Quantum measurements are intrinsically probabilistic, collapsing the state into an eigenstate of the measurement operator. QST reconstructs the full state by performing repeated measurements on an ensemble of identical states, which, together with Born's rule~\cite{Nielsen2010}, allows the reconstruction of a density matrix consistent with the observed data. It is indispensable for verifying state preparation fidelity and characterizing algorithm outputs.

Despite its importance, QST faces major challenges in practical settings, where quantum algorithms often involve processing large datasets and require dense amplitude encoding across many qubits. In such cases, the number of parameters needed to describe the quantum state grows exponentially, resulting in significant computational overhead that can offset the expected advantages of quantum algorithms~\cite{Gross2010, Riofrio2017}. This scaling issue makes full state reconstruction computationally expensive and impractical for large systems, limiting the broader application of quantum algorithms in fields such as QCFD and simulations of other complex systems.

Multiple QST techniques have been developed to address these challenges, including maximum-likelihood estimation~\cite{Banaszek1999, Shang2017}, gradient-based optimization~\cite{Ferrie2014, Bolduc2017}, machine learning approaches~\cite{Torlai2020, Ahmed2021, Koutny2022}, and other advanced schemes~\cite{Huang2020, Cotler2020, Liu2020}. More efficient methods can be established if extra restrictions are imposed on the quantum states. Examples include compressed sensing~\cite{Gross2010, Steffens2017} for low-rank density matrices and matrix product state (MPS) tomography~\cite{Cramer2010, Lanyon2017} regarding low-dimensional productive density matrices, simplifying the reconstruction of entangled systems. Overall, these methods aim to find more concise expressions of certain quantum states to shrink down the parameter space, allowing for more efficient state reconstruction in specific contexts. 
 
We present the QST with Chebyshev polynomials (QST-CP), a spectral approach that incorporates scale information through Chebyshev polynomial decomposition. 
Each Chebyshev basis function represents a scale-dependent mode and prioritizes large-scale structures over finer-scale details. Low-order modes capture the large-scale, energetically dominant features of the encoded function, while higher-order modes describe finer-scale variations. This hierarchical organization naturally supports effective truncation, as retaining only the most informative modes allows the essential physical content of the state to be preserved while mitigating the measurement overhead required for QST.

\begin{figure}
    \centering
    \includegraphics[width=0.8\linewidth]{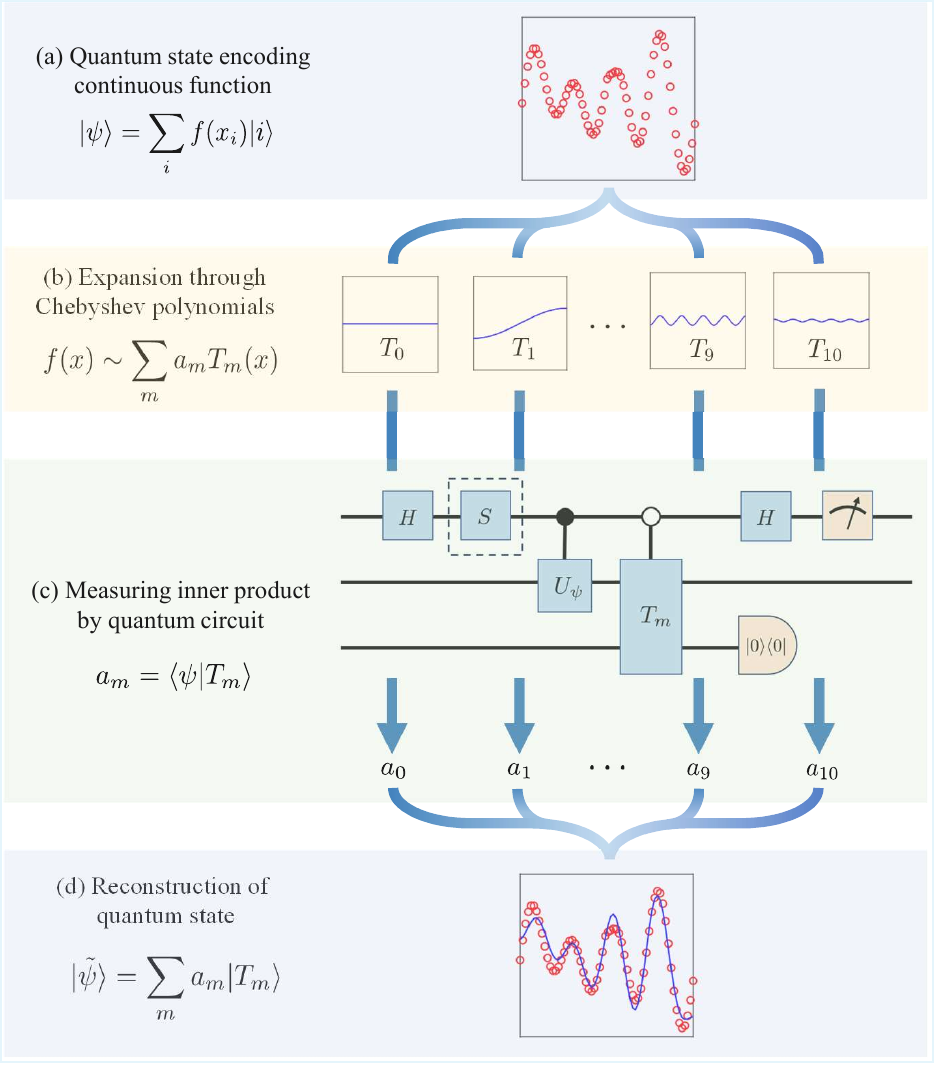}
    \caption{Overview of QST-CP. (a) A pure quantum state encoding a continuous function $f(x)$. (b) Expansion of $f(x)$ using Chebyshev polynomials, truncated to approximate the original function. (c) Measurement of expansion coefficients as inner products between the target quantum state and Chebyshev basis functions via quantum circuits. (d) Reconstruction of the truncated series, enabling approximate tomography of the quantum state.}
    \label{fig:overall}
\end{figure}

The QST-CP is applied to pure quantum states encoding continuous functions, commonly arising in solving differential equations~\cite{Kyriienko2021, Jin2023, Sarma2024}, where solutions are encoded in the amplitudes of a quantum state. By expanding these functions in terms of Chebyshev polynomials, we map the target function into a subspace spanned by a truncated Chebyshev series. 
The expansion coefficients, defined as inner products between the target function and the basis polynomials, are then measured via quantum circuits, as illustrated in \figref{fig:overall}.
Given the quantum state encoding the target function in \figref{fig:overall}(a) together with the Chebyshev polynomial expansions in \figref{fig:overall}(b), the quantum circuit shown in \figref{fig:overall}(c) is employed to measure the expansion coefficients, from which the quantum state is reconstructed, as shown in \figref{fig:overall}(d). QST-CP provides a balance between efficiency and accuracy by adjusting the truncation length of the polynomial series. Short truncations significantly improve QST efficiency and retain large-scale modes, while longer truncations capture finer details at the cost of increased measurement expense. Validations are performed by reconstructing analytic functions and complex flow fields, demonstrating the robustness and efficacy of QST-CP in quantum state reconstruction~\cite{github}. 

This paper is structured as follows. The theoretical framework of QST-CP is introduced in Section~\ref{sec:theory}, and the corresponding circuit implementation is described in Section~\ref{sec:implementation}. Numerical demonstrations are presented in Section~\ref{sec:result}, and concluding remarks are given in Section~\ref{sec:conclusion}.

\section{QST with Chebyshev polynomials}\label{sec:theory}

\subsection{QST on subspace}
For a system of $n$ qubits, the underlying Hilbert space is 
$\mathcal{H} = (\mathbb{C}^2)^{\otimes n}$ with dimension $2^n$, where $\mathbb{C}$ denotes the complex plane. 
Physically distinct pure states correspond to the complex projective space 
$\mathbb{CP}^{\,2^n - 1}$, obtained by identifying normalized vectors in $\mathcal{H}$ that differ only by a global phase.
A general quantum state is described by a density operator $\varrho \in \mathcal{B}(\mathcal{H})$, where $\mathcal{B}(\cdot)$ denotes the space of bounded linear operator.  
Density operator $\varrho$ is a $2^n \times 2^n$ Hermitian, positive semidefinite operator with unit trace, and its set forms a convex subset of $\mathbb{R}^{(2^n)^2-1}$. Hence, an $n$-qubit state is characterized by $4^n - 1$ independent real parameters.

The QST aims to reconstruct $\varrho$ by performing measurements in different bases and estimating the outcome probabilities from Born's rule. Since the number of free parameters grows exponentially with $n$, standard QST typically requires on the order of $4^n$ measurement settings, rendering it impractical for large systems.

In many applications the relevant quantum states are not arbitrary but encode structured data. Consider a continuous function $f(x_1,\ldots,x_d)$ discretized on a grid of size $2^{n_1}\times \cdots \times 2^{n_d}$ with coordinates $(x_{k_1},\ldots,x_{k_d})$. Its quantum state can be expressed as
\begin{equation}\label{eq:encoding}
    \ket{\psi}=\sum_{k_1=0}^{2^{n_1}-1}\cdots\sum_{k_d=0}^{2^{n_d}-1}
    f(x_{k_1}, \ldots, x_{k_d}) \ket{k_1}\otimes\cdots\otimes\ket{k_d}.
\end{equation}
Such state representations naturally arise in quantum algorithms for solving differential equations~\cite{Kyriienko2021, Jin2023, Sarma2024} and in QCFD, where solutions are stored in state amplitudes and efficient readout is essential. Many physically relevant functions exhibit scale-dependent behavior, allowing for the reconstruction of only a subset of amplitudes instead of performing full tomography on all of them. For instance, turbulent velocity fields follow the Kolmogorov theory~\cite{Kolmogorov1941}, where the energy spectrum decays with wavenumber and obeys a power-law distribution in the inertial range. In such cases large-scale components dominate the overall behavior. 

In classical approximation theory, continuous functions are represented using orthogonal bases. By expanding $f$ in terms of Chebyshev polynomials and restricting the state to the subspace spanned by a truncated set of basis functions, one can retain the dominant large-scale content with relatively few coefficients. Tomography then reduces to estimating these coefficients, and such parametrization substantially lowers the measurement overhead while preserving the essential physical features of the quantum state.

\subsection{Chebyshev polynomials}
Chebyshev polynomials are a family of orthogonal polynomials that play an important role in approximation theory. 
They are frequently employed to represent continuous functions, for example in numerical methods for differential equations, where physical quantities such as pressure, velocity, or temperature fields can be expanded using orthogonal polynomial bases.

The Chebyshev polynomial of degree $p$ is defined as 
$T_p(x) = \cos(p \arccos x)$ for $-1 \leq x \leq 1$, 
and it satisfies the recurrence relation~\cite{Rivlin2020}
\begin{equation}\label{eq:Chebyshev_definition}
T_{p+1}(x) = 2xT_p(x) - T_{p-1}(x),\quad p \geq 1,
\end{equation}
with initial conditions $T_0(x)=1$ and $T_1(x)=x$. 
For example, $T_2(x)=2x^2-1$, $T_3(x)=4x^3-3x$, and $T_4(x)=8x^4-8x^2+1$. 
The polynomial $T_p(x)$ has $p$ distinct zeros in $[-1,1]$, given by 
\begin{equation}\label{eq:zero}
X_{p,k} = \cos\frac{(2k+1)\pi}{2p},\quad k = 0, \cdots, p-1,
\end{equation}
which follow directly from the trigonometric definition. 
More generally, evaluating $T_p(x)$ at the zeros of $T_q(x)$ yields 
\begin{equation}\label{eq:evaluation}
T_p(X_{q,k}) = \cos\frac{(2k+1)p\pi}{2q},\quad k = 0, \cdots, p-1.
\end{equation}

Chebyshev polynomials satisfy a discrete orthogonality relation with respect to the set of zeros $\{X_{p,k}\}$ of $T_p$. 
Defining the discrete inner product as
\[
\langle f,g\rangle_p \equiv \sum_{k=0}^{p-1} f(X_{p,k}) g(X_{p,k}),
\]
the orthogonality relation reads
\begin{equation}\label{eq:orthogonal}
\langle T_s, T_t\rangle_p=\begin{cases}
0, &s\neq t,\\
p/2, &s=t\neq 0,\\
p, &s=t=0,
\end{cases}
\end{equation}
for $0\leq s,t\leq p$. 
Based on this, an orthonormal basis can be constructed as 
$\tilde{T}_{i,p}(x) \equiv T_i(x)/\sqrt{\langle T_i, T_i\rangle_p}$. 
Any function $f \in C([-1,1])$, real- or complex-valued, can then be approximated in terms of a truncated expansion
\begin{equation}\label{eq:1d_theoretical}
f(x)\sim f_P(x)=\sum_{s=0}^P a_s \tilde{T}_{s,p}(x), 
\quad a_s=\langle f,\tilde{T}_{s,p}\rangle_p,\quad 0\leq P\leq p-1.
\end{equation}
The restriction $0\leq P\leq p-1$ comes from the fact that $f_{p-1}$ already accurately reproduces values on discrete points, as $f(X_{p,k})=f_{p-1}(X_{p,k})$, and orthogonality relation fails for higher polynomial degrees. This expansion converges for piecewise continuous functions on $[-1,1]$, and a classical error bound is available~\cite{Trefethen2019}: 
if $f^{(0)},f^{(1)},\ldots,f^{(r-1)}$ are absolutely continuous and $f^{(r)}$ has bounded variation $V$ on $[-1,1]$, then for $P>r$ the truncated series $f_P$ satisfies 
$\max_{[-1,1]}|f(x)-f_P(x)|\leq 2V/(\pi r(P-r)^r)$.
Hence the expansion achieves an algebraic convergence rate $O(P^{-r})$ for functions of sufficient smoothness, enabling approximate QST using truncated series.

The orthogonality extends naturally to multivariate functions through the tensor product basis ${\tilde{T}_{s_1, p_1}(x_1) \cdots \tilde{T}_{s_d, p_d}(x_d)}$. For $f(x_1, \ldots, x_d)$ on $[-1,1]^d$, the expansion has the form
\begin{equation}\label{eq:multi_theoretical}
f(x_1, \cdots, x_d)\sim\sum_{s_1=0}^{P_1}\cdots\sum_{s_d=0}^{P_d}a_{s_1, \cdots, s_d}\tilde{T}_{s_1, p_1}(x_1)\cdots\tilde{T}_{s_d, p_d}(x_d),\quad 0\leq P_i\leq p_i-1,
\end{equation}
with expansion coefficients
\begin{equation}
\begin{aligned}
a_{s_1, \cdots, s_d}&=\langle f, \tilde{T}_{s_1, p_1}\cdots\tilde{T}_{s_d, p_d}\rangle_{p_1, \cdots, p_d}\\
&\equiv\sum_{k_1=0}^{p_1-1}\cdots\sum_{k_d=0}^{p_d-1}f(X_{p_1, k_1}, \cdots, X_{p_d, k_d})\tilde{T}_{s_1, p_1}(X_{p_1, k_1})\cdots\tilde{T}_{s_d, p_d}(X_{p_d, k_d}).
\end{aligned}
\end{equation} 
While exact function reconstruction requires the coefficients to match the number of grid points, smooth functions can be approximated using low-order terms that capture the large-scale behavior. This mathematical structure motivates our quantum approach, where we efficiently extract expansion coefficients through quantum circuits to achieve approximate QST, thereby circumventing the traditional exponential costs.

\section{Implementation of QST-CP}\label{sec:implementation}
\subsection{QST-CP for single-variable functions}
For single-variable functions, the quantum state in Eq.~(\ref{eq:encoding}) simplifies to $\ket{\psi}=\sum_{k=0}^{2^n-1}f(x_k)\ket{k}$, where $n$ is the number of qubits, $k$ labels the computational basis states, and $x_k$ denote uniformly spaced sampling points in $[-1,1]$. To enable the Chebyshev expansion, we introduce a variable substitution $x_k\rightarrow X_{2^n, k}$, defining a new target function $F(X_{2^n, k})=f(x_{k})$, The quantum state then becomes $\ket{\psi}=\sum_{k=0}^{2^n-1}F(X_{2^n, k})\ket{k}$. 

To measure $\ket{\psi}$, we introduce the normalized Chebyshev polynomials encoded by
\begin{equation}
    \ket{T_{s,n}}=\sum_{k=0}^{2^n-1}\tilde{T}_{s, 2^n}(X_{2^n, k})\ket{k}. 
\end{equation}
Thus the quantum state inner product equals the discrete polynomial inner product $\langle\cdot, \cdot\rangle_{2^n}$ with following derivation
\[\braket{T_{s,n}|\psi}=\sum_{k=0}^{2^n-1}\overline{\tilde{T}_{s, 2^n}(X_{2^n, k})}F(X_{2^n,k})=\sum_{k=0}^{2^n-1}\tilde{T}_{s, 2^n}(X_{2^n, k})F(X_{2^n,k})=\langle F, \tilde{T}_{s, 2^n}\rangle_{2^n},\]
where the second equality follows from the real-valued nature of $\tilde{T}_{s,2^n}(\cdot)$, allowing the expansion coefficients to be obtained directly via quantum state inner product measurements.

\begin{figure}
    \centering
    \includegraphics[width=0.6\linewidth]{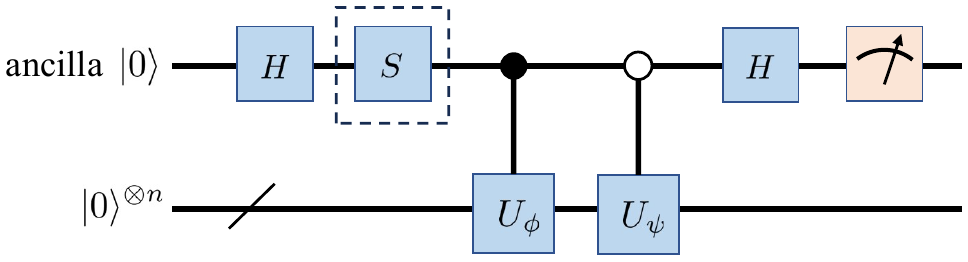}
    \caption{Circuit for measuring inner product $\langle \phi|\psi\rangle$, where $U_\phi$ and $U_\psi$ are the preparation circuit for $|\phi\rangle$ and $|\psi\rangle$. The measurement output switches between obtaining the real and imaginary parts of the inner product controlled by removal or inclusion of $S$ gate enclosed by dashed line. }
    \label{fig:Hadamard}
\end{figure}

We employ the Hadamard test~\cite{Nielsen2010} presented in \figref{fig:Hadamard} to estimate this inner product. 
By removing or including the phase gate $S$ in \figref{fig:Hadamard}, the circuit is able to measure the real or imaginary part of the inner product, respectively. Given state preparation unitaries $\ket{\phi}=U_{\phi}\ket{0}$ and $\ket{\psi}=U_{\psi}\ket{0}$, we obtain $\mathrm{Prob}(\ket{0})=(1+\mathrm{Re}\langle\phi|\psi\rangle)/2$ without the $S$ gate, and $\mathrm{Prob}(\ket{0})=(1-\mathrm{Im}\langle\phi|\psi\rangle)/2$ with the gate, where $\mathrm{Prob}(\ket{0})$ denotes the probability of obtaining output $\ket{0}$. This protocol separates estimations of both real and imaginary components of the inner product, extending QST-CP to general quantum states with complex amplitudes. 

\begin{figure}
    \centering
    \includegraphics[width=0.9\linewidth]{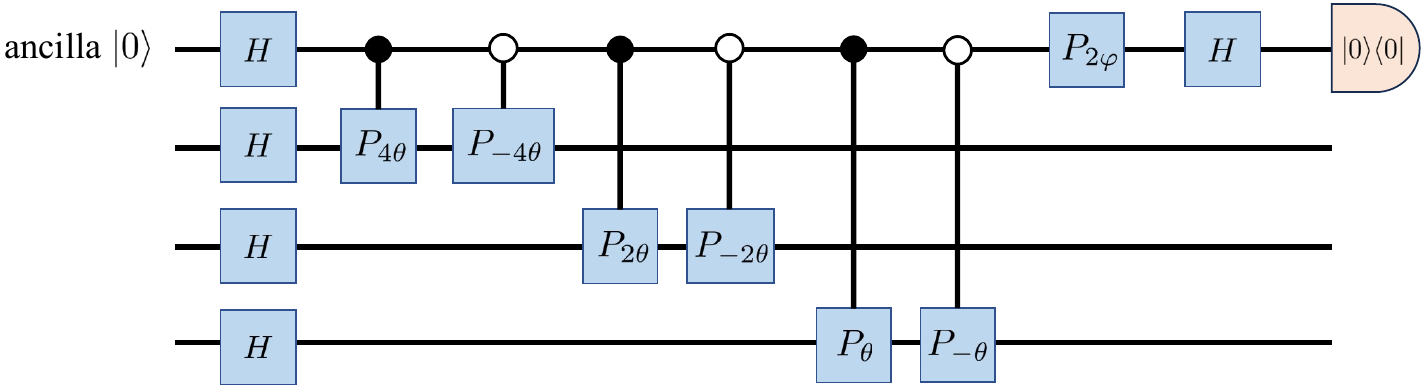}
    \caption{Preparation circuit for $|T_{s,n}\rangle$ with $n=3$, $\theta=s\pi/2^n$, and $\varphi=s\pi/2^{n+1}$. The $\ket{0}\bra{0}$ operation on the end of ancilla qubit refers to measuring the qubit so that the result is $\ket{0}$. }
    \label{fig:prep}
\end{figure}

We prepare the state $\ket{T_{s,n}}$ using the Chebyshev polynomial values
\begin{equation}\label{eq:value}
    T_s(X_{2^n,k})=\cos\frac{(2k+1)s\pi}{2^{n+1}}
\end{equation}
at the Chebyshev nodes $X_{2^n,k}$ via \eqref{eq:evaluation}, where the normalization factor for $\tilde{T}_{s,2^n}$ is omitted here. 
The corresponding quantum circuit in \figref{fig:prep} is inspired by the standard quantum Fourier transform circuit~\cite{Nielsen2010}. The controlled-gate segment generates the quantum state 
\[\ket{\psi}=\frac{1}{2^{n/2}}\left(\sum_{k=0}^{2^n-1}e^{-ik\theta}\ket{0}\otimes\ket{k}+\sum_{k=0}^{2^n-1}e^{ik\theta}\ket{1}\otimes\ket{k}\right).\]
To convert the relative phase between the $\ket{0}$ and $\ket{1}$ components into a measurable amplitude difference in the computational register, subsequent application of the $P_{2\varphi}$ gate followed by a Hadamard gate to the ancilla qubit yields the state
\[\begin{aligned}
HP_{2\varphi}\ket{\psi}&= \frac{1}{2^{n/2}}H\left(\sum_{k=0}^{2^n-1}e^{-ik\theta}\ket{0}\otimes\ket{k}+\sum_{k=0}^{2^n-1}e^{i(k\theta+2\varphi)}\ket{1}\otimes\ket{k}\right)\\
&=\frac{1}{2^{n/2}}\left(\sum_{k=0}^{2^n-1}\frac{1}{2}(e^{-ik\theta}+e^{i(k\theta+2\varphi)})\ket{0}\otimes\ket{k}+\sum_{k=0}^{2^n-1}\frac{1}{2}(e^{-ik\theta}-e^{i(k\theta+2\varphi)})\ket{1}\otimes\ket{k}\right)\\
&=\frac{1}{2^{n/2}}\left(e^{i\varphi}\ket{0}\otimes\sum_{k=0}^{2^n-1}\cos(k\theta+\varphi)\ket{k}+e^{i\varphi}\ket{1}\otimes\sum_{k=0}^{2^n-1}\sin(k\theta+\varphi)\ket{k}\right)\\
&=\frac{1}{2^{n/2}}\Bigg(e^{i\varphi}\ket{0}\otimes\sum_{k=0}^{2^n-1}\cos\frac{(2k+1)s\pi}{2^{n+1}}\ket{k}+e^{i\varphi}\ket{1}\otimes\sum_{k=0}^{2^n-1}\sin\frac{(2k+1)s\pi}{2^{n+1}}\ket{k}\Bigg)
\end{aligned}\]

with substituting $\theta=s\pi/2^n$ and $\varphi=s\pi/2^{n+1}$. By selecting the $\ket{0}$ part of ancilla qubit, we reproduce the magnitude values specified in \eqref{eq:value} up to an irrelevant global phase factor $e^{i\varphi}$. 

\begin{figure}
    \centering
    \includegraphics[width=0.7\linewidth]{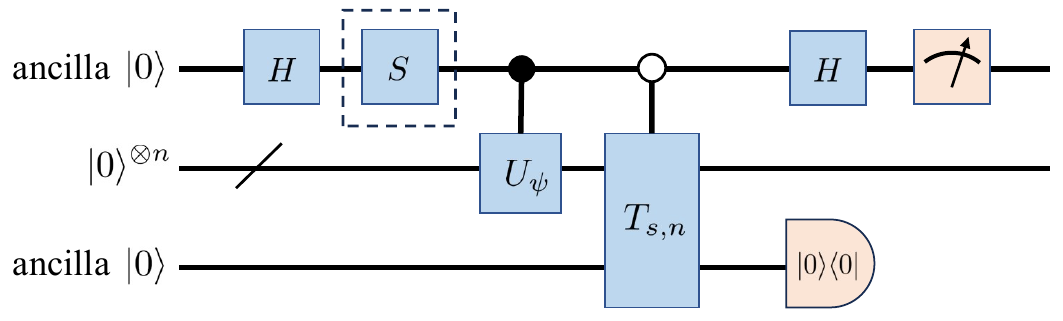}
    \caption{The complete circuit for measuring the inner product $\langle T_{s,n}|\psi\rangle$. }
    \label{fig:1dmeasure}
\end{figure}

The full circuit for inner product measurement is presented in \figref{fig:1dmeasure}. We now analyze the possibility of operation $\ket{0}\bra{0}$ outputting $\ket{0}$, which is the success probability of state preparation. First, Eqs. (\ref{eq:evaluation}) and (\ref{eq:orthogonal}) establish 
\[\langle T_s, T_s\rangle_{p}=\sum_{k=0}^{p-1}T_s(X_{p,k})^2=\sum_{k=0}^{p-1}\cos^2\frac{(2k+1)s\pi}{2p}=\frac{p}{2}.\]
Substituting $p=2^n$ into this equation yields
\[\sum_{k=0}^{2^n-1}\cos^2\frac{(2k+1)s\pi}{2^{n+1}}=\sum_{k=0}^{2^n-1}\cos^2(k\theta+\varphi)=2^{n-1}.\]
Then we have
\[\sum_{k=0}^{2^n-1}\sin^2(k\theta+\varphi)=2^n-\sum_{k=0}^{2^n-1}\cos^2(k\theta+\varphi)=2^{n-1}=\sum_{k=0}^{2^n-1}\cos^2(k\theta+\varphi).\]
This implies that the ancilla qubit used to prepare $\ket{T_{s,n}}$ in \figref{fig:prep} has an equal probability of collapsing to either $\ket{0}$ or $\ket{1}$. In the complete measurement circuit \figref{fig:1dmeasure}, the first ancilla qubit similarly exhibits equal probability of $\ket{0}$ and $\ket{1}$ after the initial Hadamard gate. Crucially, the $\ket{T_{s,n}}$ state preparation only occurs in the $\ket{0}$ branch. Consequently, the $\ket{0}\bra{0}$ measurement yields $\ket{0}$ with $\mathrm{Prob}(\ket{0})=1-0.5\times0.5=0.75$, i.e., the overall success probability for preparing $\ket{T_{s,n}}$ is $75\%$. 

Theoretically, the inner product measurement can be extended up to the $(2^n-1)$th-order polynomial, though this is computationally expensive and typically unnecessary in practice. To establish a stopping criterion for the measurement, we leverage the algebraic identity $\sum_{s=0}^{2^n-1}|a_s|^2=\sum_{k=0}^{2^n-1}|f(x_k)|^2=1$, where the first equality mirrors Parseval's theorem, and the second follows from the quantum state normalization condition. We then define the partial sum 
\begin{equation}\label{eq:partial_sum}
    A_m\equiv\sum_{s=0}^m|a_s|^2\leq 1 
\end{equation}
to quantify the fraction of the original function's information captured by the Chebyshev expansion up to order $m$. By choosing a threshold value $A_c\in(0,1)$, we stop the measurement procedure once $A_m\geq A_c$. This provides a simple and efficient stopping criterion, balancing efficiency and accuracy. 

\subsection{QST-CP for multi-variable functions}
\begin{figure}
    \centering
    \includegraphics[width=0.7\linewidth]{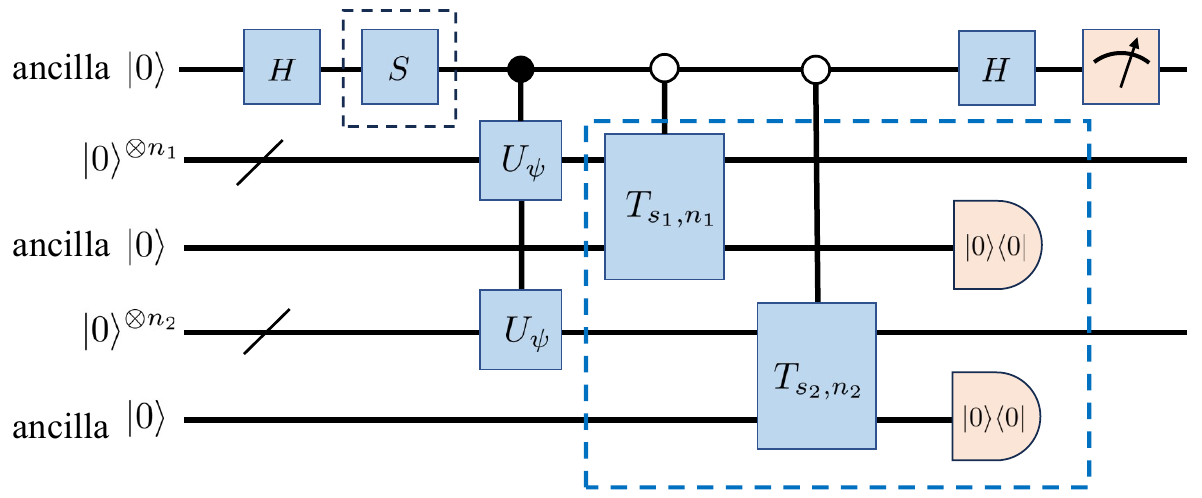}
    \caption{Example of a circuit measuring the inner product between $\ket{\psi}$ and $\ket{T_{s_1,n_1}}\otimes\ket{T_{s_2,n_2}}$. The part enclosed by the dashed blue box is the preparation circuit of the multivariate polynomial basis $\ket{T_{s_1,n_1}}\otimes\ket{T_{s_2,n_2}}$. }
    \label{fig:2d_circuit}
\end{figure}

For the multivariate case, we recall that $\{\tilde{T}_{s_1,p_1}(x_1)\cdots\tilde{T}_{s_d,p_d}(x_d)\}$ forms an orthonormal system. The corresponding quantum state admits the decomposition
\begin{equation}
\begin{aligned}
    \ket{T_{s_1, \cdots, s_n, n_1, \cdots, n_d}}&=\sum_{k_1=0}^{2^{n_1}-1}\cdots\sum_{k_d=0}^{2^{n_d}-1}\tilde{T}_{s_1, 2^{n_1}}(X_{2^{n_1},s_1})\cdots\tilde{T}_{s_d,2^{n_d}}(X_{2^{n_d},s_d})\ket{k_1}\otimes\cdots\otimes\ket{k_d}\\
    &=\ket{T_{s_1,n_1}}\otimes\cdots\otimes\ket{T_{s_d,n_d}}.
\end{aligned}
\end{equation}
This factorization infers that a multivariate Chebyshev polynomial can be prepared through parallel composition of single-variable circuits, as illustrated in Fig.~\ref{fig:2d_circuit}. 

The partial sum in \eqref{eq:partial_sum} can be generalized to multivariate cases as
\begin{equation}
    A_m\equiv \sum_{0\leq s_1+\cdots+s_d\leq m}|a_{s_1,\cdots,s_d}|^2,
\end{equation}
where the summation runs over multi-indices $(s_1, \cdots, s_n)$ with total degree not surpassing $m$. This yields a stopping criterion for multivariate measurements, where given a threshold $A_c\in(0,1)$, we measure all inner products corresponding to $m$-th order polynomials and stop upon $A_m \geq A_c$.

\subsection{Complexity analysis}
We present a complexity analysis of QST-CP for measuring $d$-variable functions encoded using $n$ qubits. The preparation circuit of $\ket{T_{s,n}}$ in \figref{fig:prep} has a depth of $O(n)$, which is efficient. 
The introduction of control qubits to the preparation circuit $U_{\psi}$ and $T_{s,n}$ in \figref{fig:1dmeasure} adds extra circuit depth, with the amount of increase depending on the specific form of the preparation circuit for target state. Deep original circuits tend to introduce deep overhead and affect actual implementation. This can be alleviated by finding equivalent and shallower preparation circuits and searching for alternative inner product measurement methods. 

The measurement repetition count mainly depends on the Chebyshev expansion order. For an $m$-th order expansion, the required number of Chebyshev bases scales as $O(m^d/d!)$. Furthermore, each basis preparation requires $d$ extra measurements, each succeeding with probability $3/4$. Thus, each basis demands $O((4/3)^d/\varepsilon)$ measurement repetitions, and the total measurement count scales as $O((4m/3)^d/(\varepsilon d!))$. The post-processing only involves summing all employed Chebyshev bases with the complexity $O(m^d/d!)$ for evaluating the magnitude of a single quantum state.

We summarize the complexities for both single- and multi-variable cases in Table \ref{tab:complexity}. 
The measurement repetition count and post-processing complexity are independent of qubit count $n$. Meanwhile, the circuit depth is linear in $n$ if the state can be prepared by a shallow circuit. This is a significant improvement over classical tomography methods requiring exponential costs. However, the complexity of QST-CP is highly dependent on the polynomial degree $m$. By controlling $m$, we can keep the overall expense manageable. 

\begin{table}
    \centering
    \begin{tabular}{cccc}
        \hline
       &Circuit depth & Repetition count & Post-processing\\
       \hline
       Single variable&$O(n)$&$O(4m/(3\varepsilon))$&$O(1)$\\
       Multi variable&$O(n)$&$O((4m/3)^d/(\varepsilon d!))$&$O(m^d/d!)$\\
       \hline
    \end{tabular}
    \caption{The circuit depth, number of measuring repetitions, and post-processing complexity of QST-CP.}
    \label{tab:complexity}
\end{table}

\section{Results}\label{sec:result}

\subsection{QST-CP of analytic functions}
We evaluate QST-CP by applying it to various functions encoded in quantum states. 
Qiskit~\cite{qiskit2024} was used to simulate the entire workflow including circuit execution and measurement on a classical computer, where the preparation circuits were implemented using its \texttt{StatePreparation} function. In all reported results, we performed $500$ measurement repetitions for each expansion coefficients. 

\begin{figure}
    \subfigure{
        \begin{minipage}{0.45\linewidth}
            \centering
            \includegraphics[width=\linewidth]{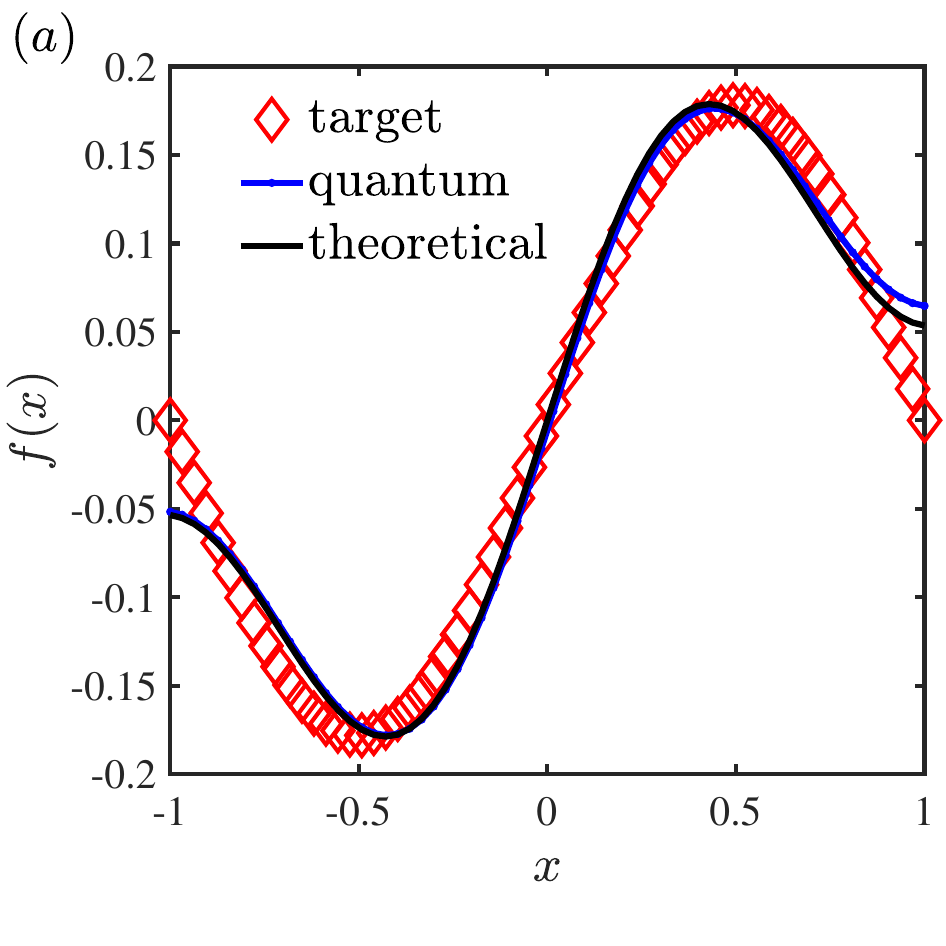}
        \end{minipage}
        \label{fig:1d_sinpix_com}
    }
    \subfigure{
        \begin{minipage}{0.45\linewidth}
            \centering
            \includegraphics[width=\linewidth]{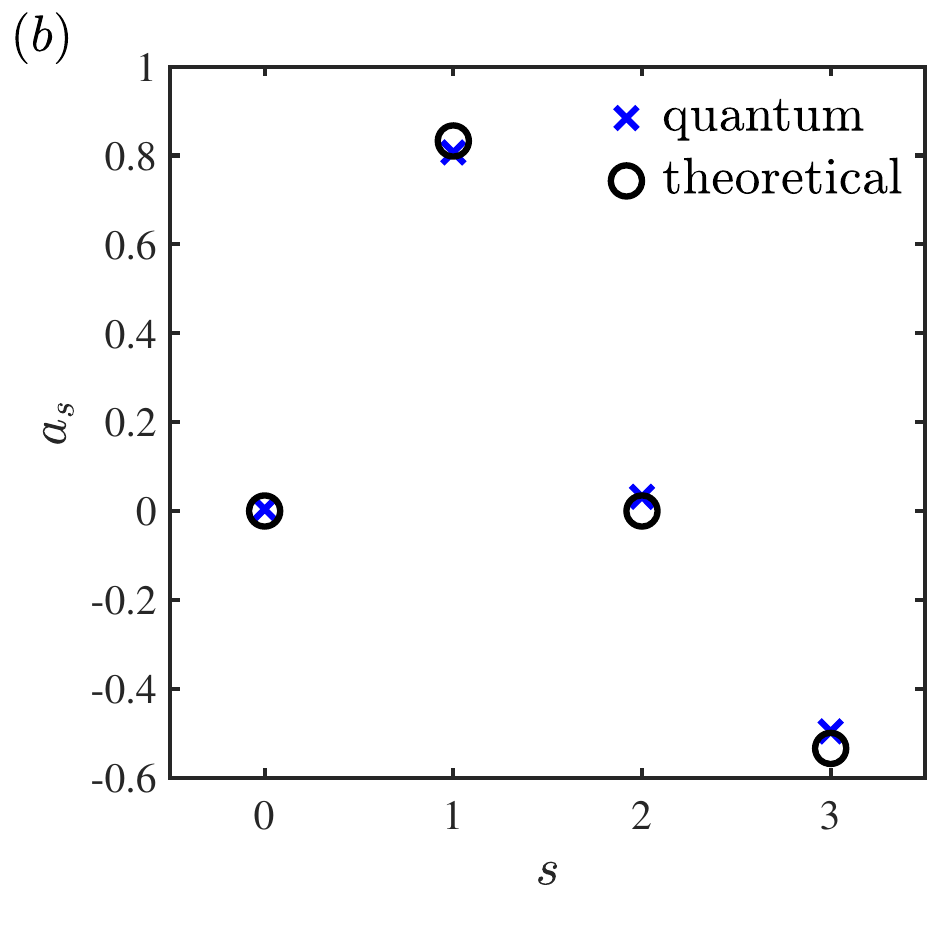}
        \end{minipage}
        \label{fig:1d_sinpix_am}
    }

    \subfigure{
        \begin{minipage}{0.45\linewidth}
            \centering
            \includegraphics[width=\linewidth]{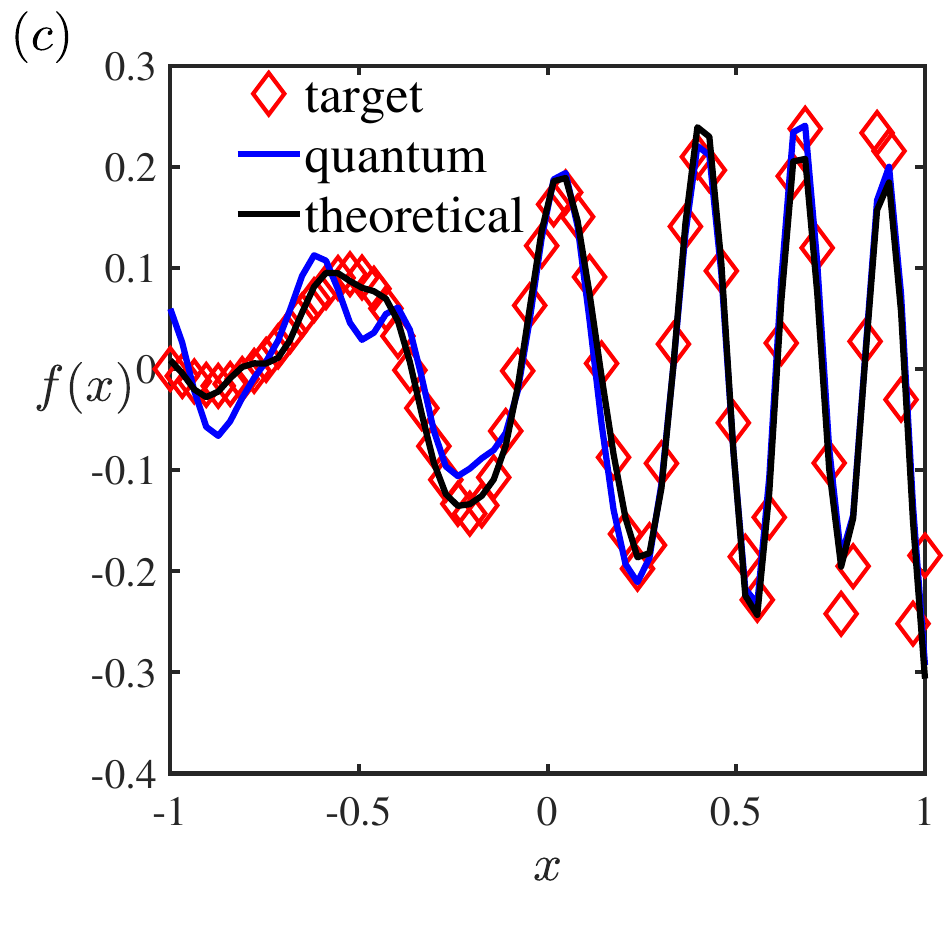}
        \end{minipage}
        \label{fig:1d_DNS_com}
    }
    \subfigure{
        \begin{minipage}{0.45\linewidth}
            \centering
            \includegraphics[width=\linewidth]{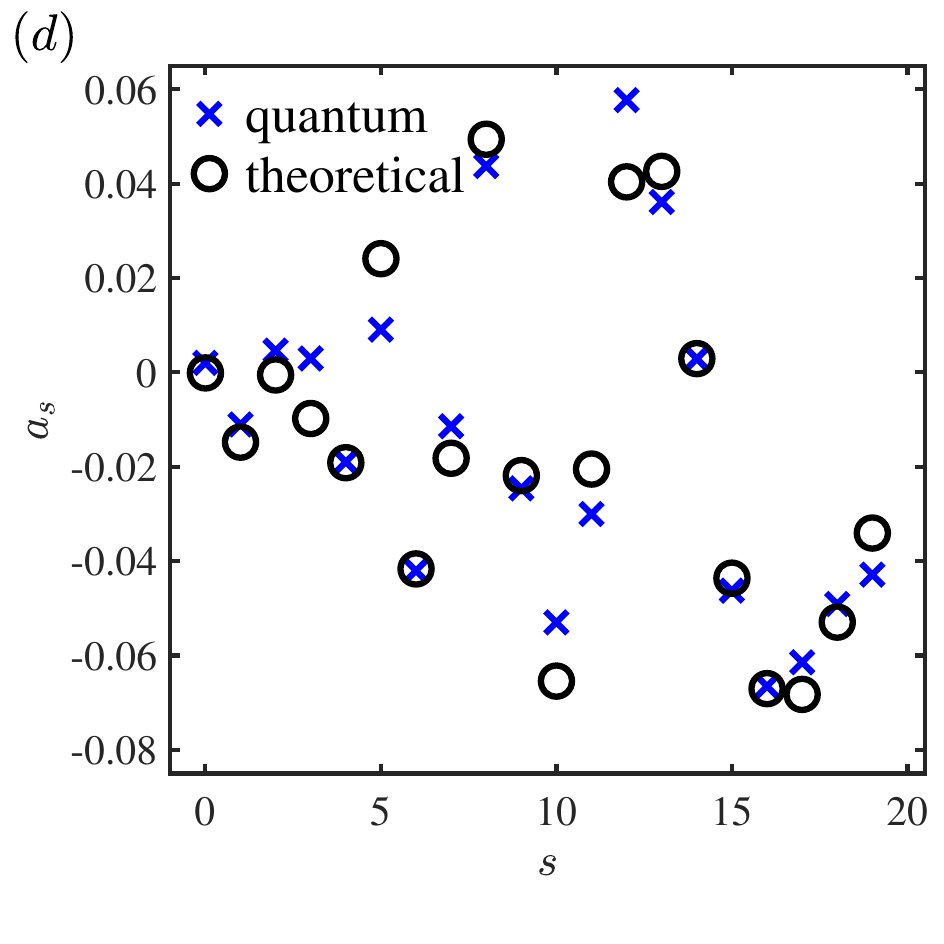}
        \end{minipage}
        \label{fig:1d_DNS_am}
    }
    \caption{Simulation results of the QST-CP method for measuring one-dimensional functions encoded with 6 qubits. Panels (a) and (b) show the single-mode function $f(x)=\sin(\pi x)$, while (c) and (d) show multi-mode function $f(x)=\log(x+1)\sin(5e^x)$. Panels (a) and (c) present the comparison among the target function, the quantum result and the theoretical result of the expansion with the same degree of polynomials, with quantum fidelity $97.74\%$ and $90.11\%$ for the single- and multi-mode cases, respectively. Panels (b) and (d) present comparison between the expansion coefficients $a_s$ acquired by quantum method and theoretical calculation.}
    \label{fig:1d_simulation}
\end{figure}

Results for measuring a single-mode function $f(x)=\sin\pi x$ and a multi-mode function $f(x)=\log(x+1)\sin(5e^x)$ with varying amplitudes, encoded using 6 qubits, are presented in \figref{fig:1d_simulation}. Note that all function values are normalized in the implementation to satisfy the normalization condition of a quantum state. 
The measurement accuracy is quantified by the quantum fidelity $\|\langle f_t|f_m\rangle\|^2$
between the target function state $\ket{f_t}$ and the measurement result state $\ket{f_m}$.  
The single- and multi-mode cases in Figs.~\ref{fig:1d_simulation}(a) and \ref{fig:1d_simulation}(c), with manageable expansion orders $m=3$ and $m=19$, yield quantum fidelities of $97.74\%$ and $90.11\%$, respectively, demonstrating that the method captures key features of analytic functions.

Figures~\ref{fig:1d_simulation}(b) and \ref{fig:1d_simulation}(d) provide a more detailed comparison for each expansion coefficients.
Most measured coefficients are close to their theoretical values calculated from \eqref{eq:1d_theoretical}, while some exhibit noticeable deviations, likely due to the limited 500 shots used for each coefficient. 
Nevertheless, the summed results still remain in good agreement with the theoretical expansion, demonstrating that QST-CP is reasonably robust against measurement-induced sampling errors, a desirable property for practical implementations on quantum hardware. 

\begin{figure}
    \subfigure{
        \begin{minipage}{0.45\linewidth}
            \centering
            \includegraphics[width=\linewidth]{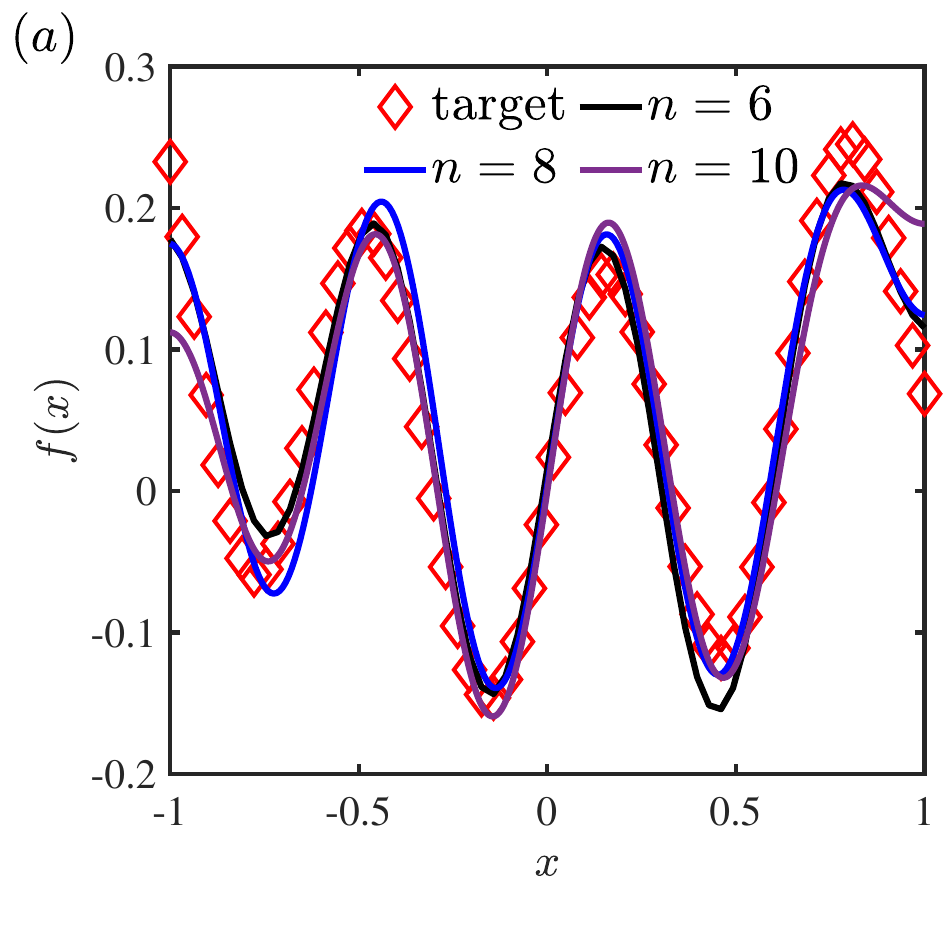}
        \end{minipage}
        \label{fig:x2sin10x_n6810_com}
    }
    \subfigure{
        \begin{minipage}{0.45\linewidth}
            \centering
            \includegraphics[width=\linewidth]{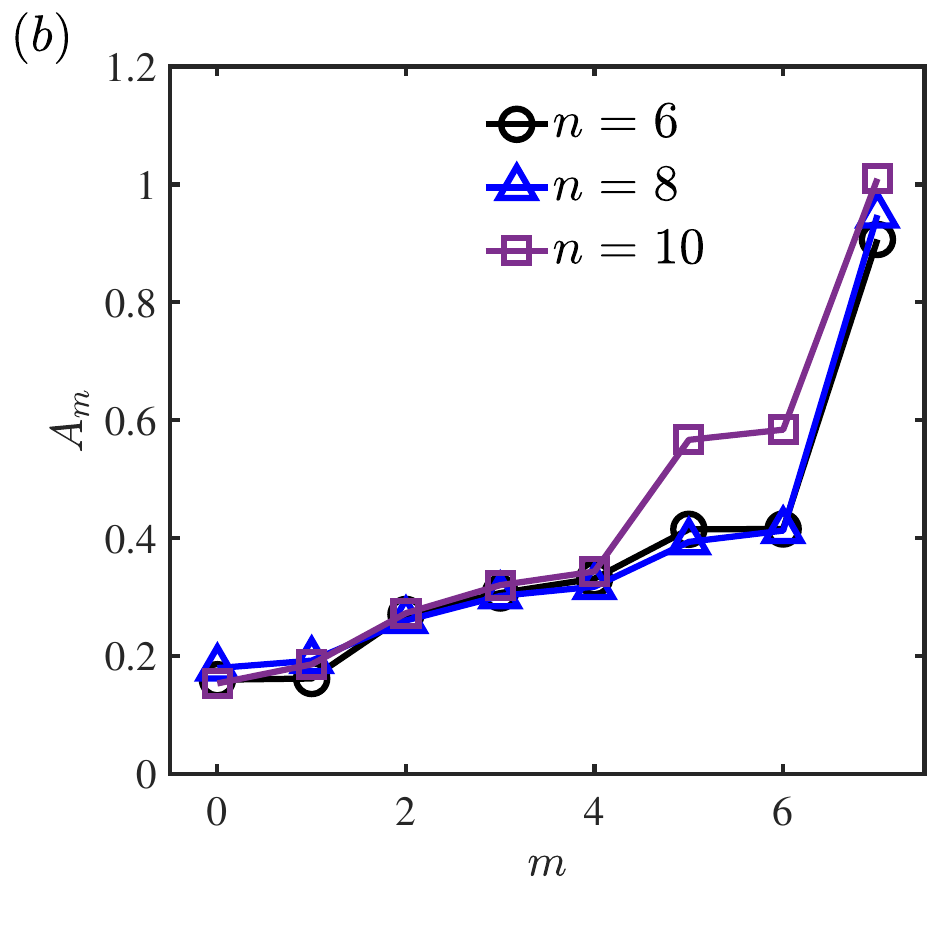}
        \end{minipage}
        \label{fig:x2sin10x_n6810_addup}
    }
    \caption{Approximate tomography result of function $f(x)=x^2+\sin(10x)$ on $[-1,1]$ encoded with  $n=6$, 8, and 10 qubits. (a) Comparison between target function and results with different $n$. For clear demonstration, the function values for $n=6$ and $8$ are normalized to collapse with $n=10$ curve. The quantum fidelities are $95.56\%$, $94.01\%$, and $93.99\%$ for these three cases. (b) The trend of $A_m$ for different $n$. The measurement ends at $m=7$ for all three cases. }
    \label{fig:x2sin10x_n6810}
\end{figure}

We further examine the dependence of QST-CP on the number of qubits by considering three quantum states with $n=6$, 8, and 10, all encoding the same function $f(x)=x^2+\sin(10x)$. Applying QST-CP with $A_c=0.85$, the measurement results in \figref{fig:x2sin10x_n6810} show good agreement with the target function, with quantum fidelity $95.56\%$, $94.01\%$, and $93.99\%$ respectively. Furthermore, the growth trends of the partial sum $A_m$ in \figref{fig:x2sin10x_n6810_addup} show that approximations with different $n$ all reach the threshold $A_c = 0.85$ at the same polynomial degree $m = 7$. Together with the observation in Table \ref{tab:complexity} that the number of measurement repetitions is independent of $n$, our approach avoids the classical exponential cost of QST for quantum states encoding simple functions.

\subsection{QST-CP of numerically simulated flow fields}
\begin{figure}
    \centering
    \includegraphics[width=\linewidth]{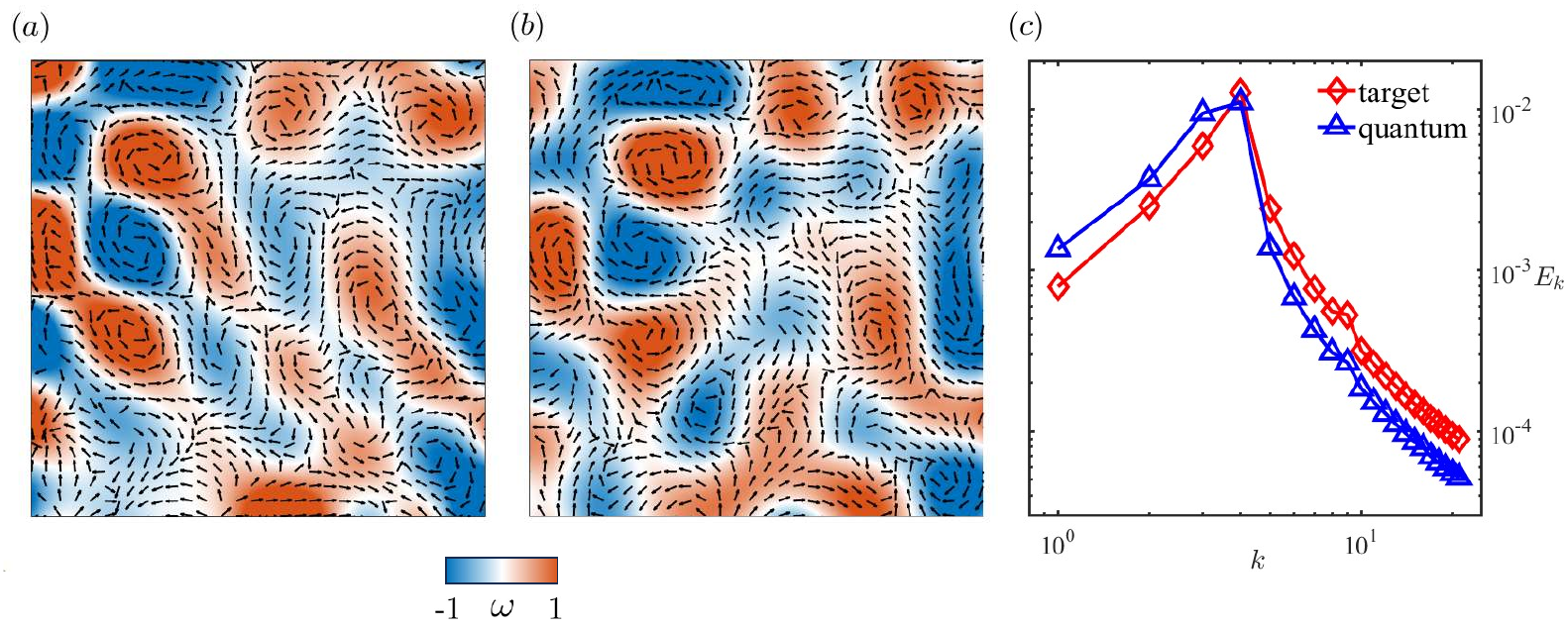}
    \caption{Comparison between (a) a snapshot of the 2D DNS velocity field (arrows) and vorticity contour and (b) the corresponding quantum measuring results on a $64 \times 64$ grid (corresponding to 12 qubits) with threshold $A_c = 0.5$ and polynomial order $m = 11$. The quantum fidelity is $78.42\%$. (c) Comparison of the energy spectra for the two velocity fields.}
    \label{fig:2d_DNS}
\end{figure}

We evaluate QST-CP on 2D vortical flows from numerical simulations, expressed as $f(x,y)=u_x(x,y)+i u_y(x,y)$ with velocity components $u_x$ and $u_y$. The flow field is generated by solving the 2D Navier--Stokes (NS) equations for homogeneous isotropic turbulence using a standard pseudo-spectral method, and the resulting data are encoded into quantum states. As a representative case, we analyze a 2D velocity field from direct numerical simulation (DNS). The simulation is calculated by pseudo-spectral method~\cite{Orszag1969} on a $[0, 2\pi]^2$ domain discretized with $1024^2$ grid points under periodic boundary conditions. The velocity field exhibits a Taylor Reynolds number of $\mathrm{Re}_{\lambda}=122.9$. The field is downsampled from $1024^2$ to $64^2$ uniform grid points, corresponding to 12 qubits. Figures~\ref{fig:2d_DNS}(a) and \ref{fig:2d_DNS}(b) compare the DNS reference field with the quantum reconstruction, where the vorticity magnitude is defined by $\omega=\partial u_y/\partial x-\partial u_x/\partial y$. With the threshold parameter set to $A_c=0.5$, the measurement terminates at polynomial order $m=11$. The reconstruction reproduces the dominant large-scale vortex structures, which is further supported by the agreement of the energy spectra shown in Fig.~\ref{fig:2d_DNS}(c), demonstrating the capability of the method to recover the major features of complex flow fields.  

\begin{figure}
    
    \includegraphics[width=\linewidth]{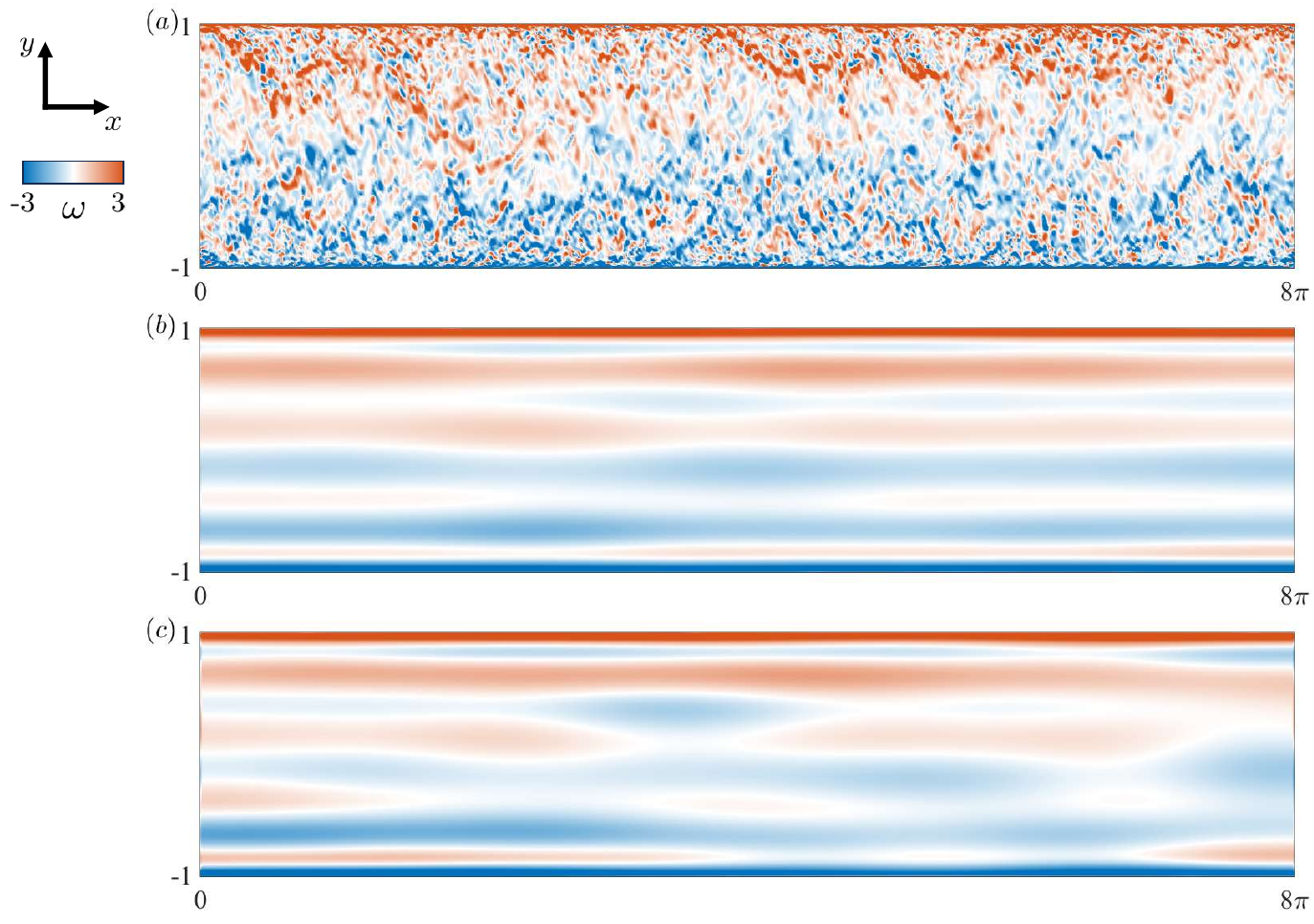}
    \caption{Result of quantum measuring a $x$-$y$ section of turbulent channel flow with $512\times 128$ uniform grid points (corresponding to $16$ qubits). (a) Target flow field from DNS data. (b) Quantum measuring result with $A_c=0.9$, leading to $m=11$. (c) Theoretical expansion result at the same polynomial order. The quantum fidelity between quantum and theoretical results is $84.33\%$. }
    \label{fig:channel_xy}
\end{figure}

To further assess the method, we consider a turbulent channel flow field from the Johns Hopkins Turbulence Database~\cite{Graham2016}, simulated in a domain of \( 8\pi \times 2 \times 3\pi \) at a friction Reynolds number of 1000. We extract $x$--$y$ and $z$--$y$ sections and resample them onto $512\times 128$ grid points, corresponding to 16 qubits, where $x$, $y$, and $z$ denote streamwise, wall-normal, and spanwise directions, respectively. Fig.~\ref{fig:channel_xy}(a) shows the target $x$--$y$ section at $z=1.5\pi$, while Figs.~\ref{fig:channel_xy}(b) and \ref{fig:channel_xy}(c) present the reconstructions, with \ref{fig:channel_xy}(b) obtained from the quantum procedure using a threshold of $A_c = 0.9$ and polynomial order $m = 11$, and \ref{fig:channel_xy}(c) from the theoretical expansion at the same order, highlighting the large-scale mode of the target flow field. The quantum reconstruction with a low polynomial degree attains a fidelity of $84.33\%$ relative to the theoretical result, effectively reproducing the dominant streamwise flow and the strong boundary vortex sheet that characterize the channel flow.

\begin{figure}
    \centering
    \includegraphics[width=\linewidth]{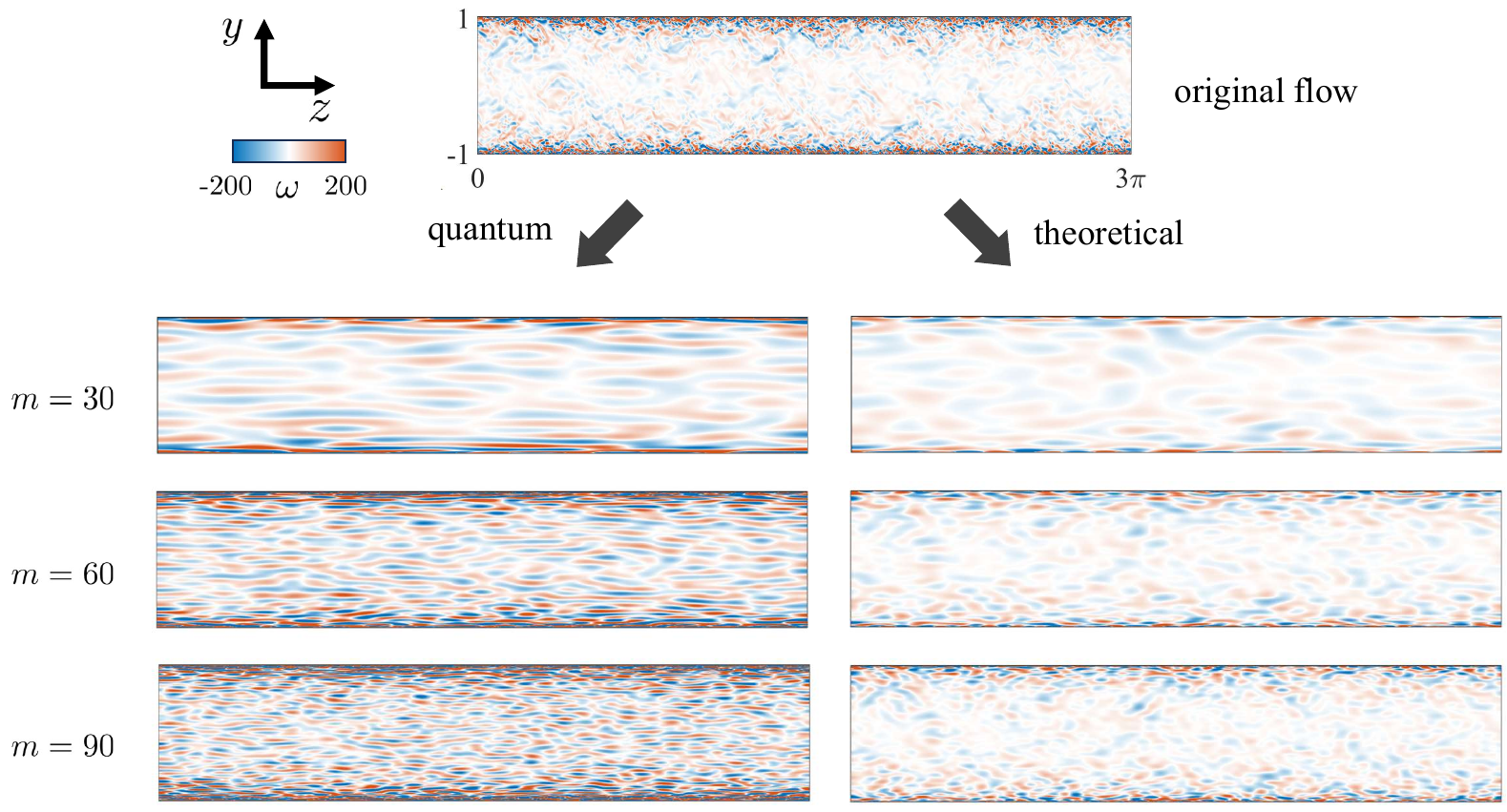}
    \caption{Result of quantum measuring a $z$-$y$ section of turbulent channel flow with $512\times 128$ uniform grid points (corresponding to $16$ qubits). Quantum results of three polynomial degrees $m=30$, 60, and 90 are provided on the left column, with theoretical results on the right column.}
    \label{fig:channel_zy}
\end{figure}

This property is further elaborated in the case of $z$-$y$ section, where $x=4\pi$ and the target function is expressed as $f(y,z)=u_z(y,z)+i u_y(y,z)$, as demonstrated in \figref{fig:channel_zy}. In the absence of the primary flow in the $x$-direction, the finer structures emerge as the dominant flow features. We provide reconstruction results for different polynomial degrees $m=30$, 60, and 90, with quantum results on the left column and theoretical results on the right. Increasing the polynomial degree progressively resolves smaller-scale features and captures richer spectral content, indicating that, by tuning the threshold value, QST-CP offers flexible control over spatial resolution to accommodate different application requirements. Lower $A_c$ values yield efficient approximate representations, whereas higher $A_c$ values enable high-resolution reconstructions that capture detailed flow characteristics.

\section{Conclusions}\label{sec:conclusion}
We propose QST-CP, a spectral method for an approximate tomography of pure quantum states encoding continuous functions, formulated in Eqs.~(\ref{eq:1d_theoretical}) and (\ref{eq:multi_theoretical}). The advantage lies in its scale-sensitive truncation, as Chebyshev modes provide a scale interpretation and low-order terms capture large-scale structures, allowing proper stopping that preserves dominant features while reducing measurement overhead. By reformulating tomography as an estimation of a set of expansion coefficients via inner product between the target quantum state and the Chebyshev bases, implemented using the circuits in Figs.~\ref{fig:Hadamard} and \ref{fig:prep}, and reconstructing a truncated series, the QST-CP method achieves polynomial scaling in circuit depth, with measurement repetitions and post-processing independent of qubit count. Furthermore, a stopping criterion based on the cumulative coefficient energy in \eqref{eq:partial_sum} establishes the accuracy-efficiency trade-off and specifies a termination rule.

We evaluate QST-CP from analytic functions to complex flow fields. For analytic functions, the method yields accurate reconstructions with modest expansion orders, regardless of the numbers of qubits used in quantum state preparation. This confirms that the quality of reconstruction remains stable as the qubit count increases, highlighting the method's scalability in addressing more complex systems. The complexity analysis in Table~\ref{tab:complexity} further reveals that the number of measurement repetitions and complexity of post-processing is independent of the qubit count, outperforming the exponential scaling typical of classical QST. For more complicated velocity fields, the method efficiently captures dominant flow structures in both isotropic turbulence and channel-flow sections with controlled expansion orders. Increasing the expansion order reveals finer-scale details, illustrating a balance between efficiency and accuracy for QST by adjusting the truncation length of the polynomial series.

The QST-CP is tailored for extracting large-scale modes from quantum states and partly addresses the output problem by enabling rapid data readout for many quantum computing applications. The algorithm can be applied without theoretical restrictions by incorporating a control qubit and the appending polynomial preparation circuit, where a shallow circuit depth ensures high efficiency. In the future work, the development of more efficient circuits for inner product measurement could reduce the reliance of overall depth and control overhead on preparation of both the target state and the polynomial basis, thereby facilitating practical deployment on quantum hardware. Moreover, integration with quantum partial differential equation solvers, where the spectral QST module is treated as a readout head for variational or block encoding solvers, could enable direct recovery of coarse solutions in quantum computations.

\section*{Acknowledgments}
This work has been supported in part by the National Natural Science Foundation of China (grant nos.~12525201, 12432010, 12588201, and 12302294), the National Key R\&D Program of China (grant No.~2023YFB4502600), and the New Cornerstone Science Foundation through the XPLORER Prize.

\section*{Code availability}
The data analysis and numerical simulation codes for this study are available for download at \url{https://github.com/YYgroup/ChebyMeasure}~\cite{github}.

\bibliographystyle{elsarticle-num-names}
\bibliography{ref}

\begin{thebibliography}{49}
\expandafter\ifx\csname natexlab\endcsname\relax\def\natexlab#1{#1}\fi
\providecommand{\url}[1]{\texttt{#1}}
\providecommand{\href}[2]{#2}
\providecommand{\path}[1]{#1}
\providecommand{\DOIprefix}{doi:}
\providecommand{\ArXivprefix}{arXiv:}
\providecommand{\URLprefix}{URL: }
\providecommand{\Pubmedprefix}{pmid:}
\providecommand{\doi}[1]{\href{http://dx.doi.org/#1}{\path{#1}}}
\providecommand{\Pubmed}[1]{\href{pmid:#1}{\path{#1}}}
\providecommand{\bibinfo}[2]{#2}
\ifx\xfnm\relax \def\xfnm[#1]{\unskip,\space#1}\fi
\bibitem[{Nielsen and Chuang(2010)}]{Nielsen2010}
\bibinfo{author}{M.~A. Nielsen}, \bibinfo{author}{I.~L. Chuang},
  \bibinfo{title}{{Quantum Computation and Quantum Information}},
  \bibinfo{publisher}{Cambridge University Press}, \bibinfo{year}{2010}.
\bibitem[{Wittek(2014)}]{Wittek2014}
\bibinfo{author}{P.~Wittek}, \bibinfo{title}{{Quantum machine learning: what
  quantum computing means to data mining}}, \bibinfo{publisher}{Academic
  Press}, \bibinfo{year}{2014}.
\bibitem[{Cao et~al.(2019)Cao, Romero, Olson, Degroote, Johnson, Kieferov{\'a},
  Kivlichan, Menke, Peropadre, Sawaya et~al.}]{Cao2019}
\bibinfo{author}{Y.~Cao}, \bibinfo{author}{J.~Romero}, \bibinfo{author}{J.~P.
  Olson}, \bibinfo{author}{M.~Degroote}, \bibinfo{author}{P.~D. Johnson},
  \bibinfo{author}{M.~Kieferov{\'a}}, \bibinfo{author}{I.~D. Kivlichan},
  \bibinfo{author}{T.~Menke}, \bibinfo{author}{B.~Peropadre},
  \bibinfo{author}{N.~P.~D. Sawaya}, et~al.,
\newblock \bibinfo{title}{{Quantum chemistry in the age of quantum computing}},
\newblock \bibinfo{journal}{Chem. Rev.} \bibinfo{volume}{119}
  (\bibinfo{year}{2019}) \bibinfo{pages}{10856--10915}.
\bibitem[{Jin et~al.(2023)Jin, Liu, and Yu}]{Jin2023}
\bibinfo{author}{S.~Jin}, \bibinfo{author}{N.~Liu}, \bibinfo{author}{Y.~Yu},
\newblock \bibinfo{title}{{Quantum simulation of partial differential
  equations: Applications and detailed analysis}},
\newblock \bibinfo{journal}{Phys. Rev. A} \bibinfo{volume}{108}
  (\bibinfo{year}{2023}) \bibinfo{pages}{032603}.
\bibitem[{Meng et~al.(2024)Meng, Zhong, Xu, Wang, Chen, Jin, Zhu, Gao, Wu,
  Zhang, Wang, Zou, Zhang, Cui, Shen, Bao, Zhu, Tan, Li, Zhang, Xiong, Li, Guo,
  Wang, Song, Wang, and Yang}]{Meng2024}
\bibinfo{author}{Z.~Meng}, \bibinfo{author}{J.~Zhong}, \bibinfo{author}{S.~Xu},
  \bibinfo{author}{K.~Wang}, \bibinfo{author}{J.~Chen},
  \bibinfo{author}{F.~Jin}, \bibinfo{author}{X.~Zhu}, \bibinfo{author}{Y.~Gao},
  \bibinfo{author}{Y.~Wu}, \bibinfo{author}{C.~Zhang},
  \bibinfo{author}{N.~Wang}, \bibinfo{author}{Y.~Zou},
  \bibinfo{author}{A.~Zhang}, \bibinfo{author}{Z.~Cui},
  \bibinfo{author}{F.~Shen}, \bibinfo{author}{Z.~Bao},
  \bibinfo{author}{Z.~Zhu}, \bibinfo{author}{Z.~Tan}, \bibinfo{author}{T.~Li},
  \bibinfo{author}{P.~Zhang}, \bibinfo{author}{S.~Xiong},
  \bibinfo{author}{H.~Li}, \bibinfo{author}{Q.~Guo}, \bibinfo{author}{Z.~Wang},
  \bibinfo{author}{C.~Song}, \bibinfo{author}{H.~Wang},
  \bibinfo{author}{Y.~Yang},
\newblock \bibinfo{title}{Simulating unsteady fluid flows on a superconducting
  quantum processor},
\newblock \bibinfo{journal}{Commun. Phys.} \bibinfo{volume}{7}
  (\bibinfo{year}{2024}) \bibinfo{pages}{349}.
\bibitem[{Xiao et~al.(2024)Xiao, Yang, Shu, Chew, Khoo, Cui, and
  Liu}]{Xiao2024}
\bibinfo{author}{Y.~Xiao}, \bibinfo{author}{L.~M. Yang},
  \bibinfo{author}{C.~Shu}, \bibinfo{author}{S.~C. Chew},
  \bibinfo{author}{B.~C. Khoo}, \bibinfo{author}{Y.~D. Cui},
  \bibinfo{author}{Y.~Y. Liu},
\newblock \bibinfo{title}{Physics-informed quantum neural network for solving
  forward and inverse problems of partial differential equations},
\newblock \bibinfo{journal}{Phys. Fluids} \bibinfo{volume}{36}
  (\bibinfo{year}{2024}) \bibinfo{pages}{9}.
\bibitem[{Succi et~al.(2023)Succi, Itani, Sreenivasan, and Steijl}]{Succi2023}
\bibinfo{author}{S.~Succi}, \bibinfo{author}{W.~Itani},
  \bibinfo{author}{K.~Sreenivasan}, \bibinfo{author}{R.~Steijl},
\newblock \bibinfo{title}{{Quantum computing for fluids: Where do we stand?}},
\newblock \bibinfo{journal}{Europhys. Lett.} \bibinfo{volume}{144}
  (\bibinfo{year}{2023}) \bibinfo{pages}{10001}.
\bibitem[{Bharadwaj and Sreenivasan(2025)}]{Bharadwaj2025}
\bibinfo{author}{S.~S. Bharadwaj}, \bibinfo{author}{K.~R. Sreenivasan},
\newblock \bibinfo{title}{{Towards simulating fluid flows with quantum
  computing}},
\newblock \bibinfo{journal}{S{\=a}dhan{\=a}} \bibinfo{volume}{50}
  (\bibinfo{year}{2025}) \bibinfo{pages}{1--19}.
\bibitem[{Tennie et~al.(2025)Tennie, Laizet, Lloyd, and Magri}]{Tennie2025}
\bibinfo{author}{F.~Tennie}, \bibinfo{author}{S.~Laizet},
  \bibinfo{author}{S.~Lloyd}, \bibinfo{author}{L.~Magri},
\newblock \bibinfo{title}{{Quantum computing for nonlinear differential
  equations and turbulence}},
\newblock \bibinfo{journal}{Nat. Rev. Phys.} \bibinfo{volume}{7}
  (\bibinfo{year}{2025}) \bibinfo{pages}{220--230}.
\bibitem[{Meng et~al.(2025)Meng, Song, and Yang}]{Meng2025}
\bibinfo{author}{Z.~Y. Meng}, \bibinfo{author}{C.~Song},
  \bibinfo{author}{Y.~Yang},
\newblock \bibinfo{title}{{Challenges of simulating fluid flows on near-term
  quantum computer}},
\newblock \bibinfo{journal}{Sci. China-Phys. Mech. Astron.}
  \bibinfo{volume}{68} (\bibinfo{year}{2025}) \bibinfo{pages}{104705}.
\bibitem[{Steijl and Barakos(2018)}]{Steijl2018}
\bibinfo{author}{R.~Steijl}, \bibinfo{author}{G.~N. Barakos},
\newblock \bibinfo{title}{{Parallel evaluation of quantum algorithms for
  computational fluid dynamics}},
\newblock \bibinfo{journal}{Comput. Fluids} \bibinfo{volume}{173}
  (\bibinfo{year}{2018}) \bibinfo{pages}{22--28}.
\bibitem[{Gaitan(2020)}]{Gaitan2020}
\bibinfo{author}{F.~Gaitan},
\newblock \bibinfo{title}{{Finding flows of a Navier-Stokes fluid through
  quantum computing}},
\newblock \bibinfo{journal}{NPJ Quantum Inf.} \bibinfo{volume}{6}
  (\bibinfo{year}{2020}) \bibinfo{pages}{61}.
\bibitem[{Lubasch et~al.(2020)Lubasch, Joo, Moinier, Kiffner, and
  Jaksch}]{Lubasch2020}
\bibinfo{author}{M.~Lubasch}, \bibinfo{author}{J.~Joo},
  \bibinfo{author}{P.~Moinier}, \bibinfo{author}{M.~Kiffner},
  \bibinfo{author}{D.~Jaksch},
\newblock \bibinfo{title}{{Variational quantum algorithms for nonlinear
  problems}},
\newblock \bibinfo{journal}{Phys. Rev. A} \bibinfo{volume}{101}
  (\bibinfo{year}{2020}) \bibinfo{pages}{010301}.
\bibitem[{Gourianov et~al.(2022)Gourianov, Lubasch, Dolgov, van~den Berg,
  Babaee, Givi, Kiffner, and Jaksch}]{Gourianov2022}
\bibinfo{author}{N.~Gourianov}, \bibinfo{author}{M.~Lubasch},
  \bibinfo{author}{S.~Dolgov}, \bibinfo{author}{Q.~Y. van~den Berg},
  \bibinfo{author}{H.~Babaee}, \bibinfo{author}{P.~Givi},
  \bibinfo{author}{M.~Kiffner}, \bibinfo{author}{D.~Jaksch},
\newblock \bibinfo{title}{{A quantum-inspired approach to exploit turbulence
  structures}},
\newblock \bibinfo{journal}{Nat. Comput. Sci.} \bibinfo{volume}{2}
  (\bibinfo{year}{2022}) \bibinfo{pages}{30--37}.
\bibitem[{Pfeffer et~al.(2022)Pfeffer, Heyder, and Schumacher}]{Pfeffer2022}
\bibinfo{author}{P.~Pfeffer}, \bibinfo{author}{F.~Heyder},
  \bibinfo{author}{J.~Schumacher},
\newblock \bibinfo{title}{{Hybrid quantum-classical reservoir computing of
  thermal convection flow}},
\newblock \bibinfo{journal}{Phys. Rev. Res.} \bibinfo{volume}{4}
  (\bibinfo{year}{2022}) \bibinfo{pages}{033176}.
\bibitem[{Zylberman et~al.(2022)Zylberman, Molfetta, Brachet, Loureiro, and
  Debbasch}]{Zylberman2022}
\bibinfo{author}{J.~Zylberman}, \bibinfo{author}{G.~D. Molfetta},
  \bibinfo{author}{M.~Brachet}, \bibinfo{author}{N.~F. Loureiro},
  \bibinfo{author}{F.~Debbasch},
\newblock \bibinfo{title}{{Quantum simulations of hydrodynamics via the
  Madelung transformation}},
\newblock \bibinfo{journal}{Phys. Rev. A} \bibinfo{volume}{106}
  (\bibinfo{year}{2022}) \bibinfo{pages}{032408}.
\bibitem[{Fukagata(2022)}]{Fukagata2022}
\bibinfo{author}{K.~Fukagata},
\newblock \bibinfo{title}{{Towards quantum computing of turbulence}},
\newblock \bibinfo{journal}{Nat. Comput. Sci.} \bibinfo{volume}{2}
  (\bibinfo{year}{2022}) \bibinfo{pages}{68--69}.
\bibitem[{Meng and Yang(2023)}]{Meng2023}
\bibinfo{author}{Z.~Y. Meng}, \bibinfo{author}{Y.~Yang},
\newblock \bibinfo{title}{{Quantum computing of fluid dynamics using the
  hydrodynamic Schr\"{o}dinger equation}},
\newblock \bibinfo{journal}{Phys. Rev. Research} \bibinfo{volume}{5}
  (\bibinfo{year}{2023}) \bibinfo{pages}{033182}.
\bibitem[{Meng and Yang(2024)}]{Meng2024quantum}
\bibinfo{author}{Z.~Y. Meng}, \bibinfo{author}{Y.~Yang},
\newblock \bibinfo{title}{{Quantum spin representation for the Navier-Stokes
  equation}},
\newblock \bibinfo{journal}{Phys. Rev. Research} \bibinfo{volume}{6}
  (\bibinfo{year}{2024}) \bibinfo{pages}{043130}.
\bibitem[{Jaksch et~al.(2023)Jaksch, Givi, Daley, and Rung}]{Jaksch2023}
\bibinfo{author}{D.~Jaksch}, \bibinfo{author}{P.~Givi}, \bibinfo{author}{A.~J.
  Daley}, \bibinfo{author}{T.~Rung},
\newblock \bibinfo{title}{{Variational quantum algorithms for computational
  fluid dynamics}},
\newblock \bibinfo{journal}{AIAA J.} \bibinfo{volume}{61}
  (\bibinfo{year}{2023}) \bibinfo{pages}{1885--1894}.
\bibitem[{Liu et~al.(2024)Liu, Chen, Shu, Rebentrost, Liu, Chew, Khoo, and
  Cui}]{Liu2024variational}
\bibinfo{author}{Y.~Liu}, \bibinfo{author}{Z.~Chen}, \bibinfo{author}{C.~Shu},
  \bibinfo{author}{P.~Rebentrost}, \bibinfo{author}{Y.~Liu},
  \bibinfo{author}{S.~Chew}, \bibinfo{author}{B.~Khoo},
  \bibinfo{author}{Y.~Cui},
\newblock \bibinfo{title}{{A variational quantum algorithm-based numerical
  method for solving potential and Stokes flows}},
\newblock \bibinfo{journal}{Ocean Eng.} \bibinfo{volume}{292}
  (\bibinfo{year}{2024}) \bibinfo{pages}{116494}.
\bibitem[{Chen et~al.(2024)Chen, Ma, Ye, Xu, Bai, Zhou, Tan, Zhuang, Xu, and
  Wang}]{Chen2024}
\bibinfo{author}{Z.~Y. Chen}, \bibinfo{author}{T.~Y. Ma},
  \bibinfo{author}{C.~C. Ye}, \bibinfo{author}{L.~Xu},
  \bibinfo{author}{W.~Bai}, \bibinfo{author}{L.~Zhou}, \bibinfo{author}{M.~Y.
  Tan}, \bibinfo{author}{X.~N. Zhuang}, \bibinfo{author}{X.~F. Xu},
  \bibinfo{author}{Y.~J. Wang},
\newblock \bibinfo{title}{{Enabling large-scale and high-precision fluid
  simulations on near-term quantum computers}},
\newblock \bibinfo{journal}{Comput. Methods Appl. Mech. Eng.}
  \bibinfo{volume}{432} (\bibinfo{year}{2024}) \bibinfo{pages}{117428}.
\bibitem[{Bharadwaj and Sreenivasan(2023)}]{Bharadwaj2023}
\bibinfo{author}{S.~S. Bharadwaj}, \bibinfo{author}{K.~R. Sreenivasan},
\newblock \bibinfo{title}{{Hybrid quantum algorithms for flow problems}},
\newblock \bibinfo{journal}{Proc. Natl. Acad. Sci.} \bibinfo{volume}{120}
  (\bibinfo{year}{2023}) \bibinfo{pages}{e2311014120}.
\bibitem[{Itani et~al.(2024)Itani, Sreenivasan, and Succi}]{Itani2024}
\bibinfo{author}{W.~Itani}, \bibinfo{author}{K.~R. Sreenivasan},
  \bibinfo{author}{S.~Succi},
\newblock \bibinfo{title}{{Quantum algorithm for lattice Boltzmann (QALB)
  simulation of incompressible fluids with a nonlinear collision term}},
\newblock \bibinfo{journal}{Phys. Fluids} \bibinfo{volume}{36}
  (\bibinfo{year}{2024}) \bibinfo{pages}{017112}.
\bibitem[{Wang et~al.(2025)Wang, Meng, Zhao, and Yang}]{Wang2025}
\bibinfo{author}{B.~Y. Wang}, \bibinfo{author}{Z.~Y. Meng},
  \bibinfo{author}{Y.~M. Zhao}, \bibinfo{author}{Y.~Yang},
\newblock \bibinfo{title}{{Quantum lattice Boltzmann method for simulating
  nonlinear fluid dynamics}},
\newblock \bibinfo{journal}{arXiv:2502.16568}  (\bibinfo{year}{2025}).
\bibitem[{Gross et~al.(2010)Gross, Liu, Flammia, Becker, and
  Eisert}]{Gross2010}
\bibinfo{author}{D.~Gross}, \bibinfo{author}{Y.~K. Liu}, \bibinfo{author}{S.~T.
  Flammia}, \bibinfo{author}{S.~Becker}, \bibinfo{author}{J.~Eisert},
\newblock \bibinfo{title}{{Quantum state tomography via compressed sensing}},
\newblock \bibinfo{journal}{Phys. Rev. Lett.} \bibinfo{volume}{105}
  (\bibinfo{year}{2010}) \bibinfo{pages}{150401}.
\bibitem[{Riofrio et~al.(2017)Riofrio, Gross, Flammia, Monz, Nigg, Blatt, and
  Eisert}]{Riofrio2017}
\bibinfo{author}{C.~A. Riofrio}, \bibinfo{author}{D.~Gross},
  \bibinfo{author}{S.~T. Flammia}, \bibinfo{author}{T.~Monz},
  \bibinfo{author}{D.~Nigg}, \bibinfo{author}{R.~Blatt},
  \bibinfo{author}{J.~Eisert},
\newblock \bibinfo{title}{{Experimental quantum compressed sensing for a
  seven-qubit system}},
\newblock \bibinfo{journal}{Nature Commun.} \bibinfo{volume}{8}
  (\bibinfo{year}{2017}) \bibinfo{pages}{15305}.
\bibitem[{Banaszek et~al.(1999)Banaszek, D'ariano, Paris, and
  Sacchi}]{Banaszek1999}
\bibinfo{author}{K.~Banaszek}, \bibinfo{author}{G.~M. D'ariano},
  \bibinfo{author}{M.~G.~A. Paris}, \bibinfo{author}{M.~F. Sacchi},
\newblock \bibinfo{title}{{Maximum-likelihood estimation of the density
  matrix}},
\newblock \bibinfo{journal}{Phys. Rev. A} \bibinfo{volume}{61}
  (\bibinfo{year}{1999}) \bibinfo{pages}{010304}.
\bibitem[{Shang et~al.(2017)Shang, Zhang, and Ng}]{Shang2017}
\bibinfo{author}{J.~W. Shang}, \bibinfo{author}{Z.~Y. Zhang},
  \bibinfo{author}{H.~K. Ng},
\newblock \bibinfo{title}{{Superfast maximum-likelihood reconstruction for
  quantum tomography}},
\newblock \bibinfo{journal}{Phys. Rev. A} \bibinfo{volume}{95}
  (\bibinfo{year}{2017}) \bibinfo{pages}{062336}.
\bibitem[{Ferrie(2014)}]{Ferrie2014}
\bibinfo{author}{C.~Ferrie},
\newblock \bibinfo{title}{{Self-guided quantum tomography}},
\newblock \bibinfo{journal}{Phys. Rev. Lett.} \bibinfo{volume}{113}
  (\bibinfo{year}{2014}) \bibinfo{pages}{190404}.
\bibitem[{Bolduc et~al.(2017)Bolduc, Knee, Gauger, and Leach}]{Bolduc2017}
\bibinfo{author}{E.~Bolduc}, \bibinfo{author}{G.~C. Knee},
  \bibinfo{author}{E.~M. Gauger}, \bibinfo{author}{J.~Leach},
\newblock \bibinfo{title}{{Projected gradient descent algorithms for quantum
  state tomography}},
\newblock \bibinfo{journal}{npj Quantum Information} \bibinfo{volume}{3}
  (\bibinfo{year}{2017}) \bibinfo{pages}{44}.
\bibitem[{Torlai and Melko(2020)}]{Torlai2020}
\bibinfo{author}{G.~Torlai}, \bibinfo{author}{R.~G. Melko},
\newblock \bibinfo{title}{{Machine-learning quantum states in the NISQ era}},
\newblock \bibinfo{journal}{Annu. Rev. Condens. Matter Phys.}
  \bibinfo{volume}{11} (\bibinfo{year}{2020}) \bibinfo{pages}{325--344}.
\bibitem[{Ahmed et~al.(2021)Ahmed, Sanchez, Nori, and Kockum}]{Ahmed2021}
\bibinfo{author}{S.~Ahmed}, \bibinfo{author}{M.~C. Sanchez},
  \bibinfo{author}{F.~Nori}, \bibinfo{author}{A.~F. Kockum},
\newblock \bibinfo{title}{{Quantum state tomography with conditional generative
  adversarial networks}},
\newblock \bibinfo{journal}{Phys. Rev. Lett.} \bibinfo{volume}{127}
  (\bibinfo{year}{2021}) \bibinfo{pages}{140502}.
\bibitem[{Koutn{\`y} et~al.(2022)Koutn{\`y}, Motka, Hradil,
  {\v{R}}eh{\'a}{\v{c}}ek, and Soto}]{Koutny2022}
\bibinfo{author}{D.~Koutn{\`y}}, \bibinfo{author}{L.~Motka},
  \bibinfo{author}{Z.~Hradil}, \bibinfo{author}{J.~{\v{R}}eh{\'a}{\v{c}}ek},
  \bibinfo{author}{L.~L.~S. Soto},
\newblock \bibinfo{title}{{Neural-network quantum state tomography}},
\newblock \bibinfo{journal}{Phys. Rev. A} \bibinfo{volume}{106}
  (\bibinfo{year}{2022}) \bibinfo{pages}{012409}.
\bibitem[{Huang et~al.(2020)Huang, Kueng, and Preskill}]{Huang2020}
\bibinfo{author}{H.~Y. Huang}, \bibinfo{author}{R.~Kueng},
  \bibinfo{author}{J.~Preskill},
\newblock \bibinfo{title}{{Predicting many properties of a quantum system from
  very few measurements}},
\newblock \bibinfo{journal}{Nature Phys.} \bibinfo{volume}{16}
  (\bibinfo{year}{2020}) \bibinfo{pages}{1050--1057}.
\bibitem[{Cotler and Wilczek(2020)}]{Cotler2020}
\bibinfo{author}{J.~Cotler}, \bibinfo{author}{F.~Wilczek},
\newblock \bibinfo{title}{{Quantum overlapping tomography}},
\newblock \bibinfo{journal}{Phys. Rev. Lett.} \bibinfo{volume}{124}
  (\bibinfo{year}{2020}) \bibinfo{pages}{100401}.
\bibitem[{Liu et~al.(2020)Liu, Wang, Xue, Huang, Fu, Qiang, Xu, Huang, Deng,
  Guo et~al.}]{Liu2020}
\bibinfo{author}{Y.~Liu}, \bibinfo{author}{D.~Y. Wang}, \bibinfo{author}{S.~C.
  Xue}, \bibinfo{author}{A.~Q. Huang}, \bibinfo{author}{X.~Fu},
  \bibinfo{author}{X.~G. Qiang}, \bibinfo{author}{P.~Xu},
  \bibinfo{author}{H.~L. Huang}, \bibinfo{author}{M.~T. Deng},
  \bibinfo{author}{C.~Guo}, et~al.,
\newblock \bibinfo{title}{{Variational quantum circuits for quantum state
  tomography}},
\newblock \bibinfo{journal}{Phys. Rev. A} \bibinfo{volume}{101}
  (\bibinfo{year}{2020}) \bibinfo{pages}{052316}.
\bibitem[{Steffens et~al.(2017)Steffens, Riofr{\'\i}o, McCutcheon, Roth, Bell,
  McMillan, Tame, Rarity, and Eisert}]{Steffens2017}
\bibinfo{author}{A.~Steffens}, \bibinfo{author}{C.~A. Riofr{\'\i}o},
  \bibinfo{author}{W.~McCutcheon}, \bibinfo{author}{I.~Roth},
  \bibinfo{author}{B.~A. Bell}, \bibinfo{author}{A.~McMillan},
  \bibinfo{author}{M.~S. Tame}, \bibinfo{author}{J.~G. Rarity},
  \bibinfo{author}{J.~Eisert},
\newblock \bibinfo{title}{{Experimentally exploring compressed sensing quantum
  tomography}},
\newblock \bibinfo{journal}{Quantum Sci. Technol.} \bibinfo{volume}{2}
  (\bibinfo{year}{2017}) \bibinfo{pages}{025005}.
\bibitem[{Cramer et~al.(2010)Cramer, Plenio, Flammia, Somma, Gross, Bartlett,
  Cardinal, Poulin, and Liu}]{Cramer2010}
\bibinfo{author}{M.~Cramer}, \bibinfo{author}{M.~B. Plenio},
  \bibinfo{author}{S.~T. Flammia}, \bibinfo{author}{R.~Somma},
  \bibinfo{author}{D.~Gross}, \bibinfo{author}{S.~D. Bartlett},
  \bibinfo{author}{O.~L. Cardinal}, \bibinfo{author}{D.~Poulin},
  \bibinfo{author}{Y.~K. Liu},
\newblock \bibinfo{title}{{Efficient quantum state tomography}},
\newblock \bibinfo{journal}{Nat. Commun.} \bibinfo{volume}{1}
  (\bibinfo{year}{2010}) \bibinfo{pages}{149}.
\bibitem[{Lanyon et~al.(2017)Lanyon, Maier, Holz{\"a}pfel, Baumgratz, Hempel,
  Jurcevic, Dhand, Buyskikh, Daley, Cramer et~al.}]{Lanyon2017}
\bibinfo{author}{B.~P. Lanyon}, \bibinfo{author}{C.~Maier},
  \bibinfo{author}{M.~Holz{\"a}pfel}, \bibinfo{author}{T.~Baumgratz},
  \bibinfo{author}{C.~Hempel}, \bibinfo{author}{P.~Jurcevic},
  \bibinfo{author}{I.~Dhand}, \bibinfo{author}{A.~S. Buyskikh},
  \bibinfo{author}{A.~J. Daley}, \bibinfo{author}{M.~Cramer}, et~al.,
\newblock \bibinfo{title}{{Efficient tomography of a quantum many-body
  system}},
\newblock \bibinfo{journal}{Nature Phys.} \bibinfo{volume}{13}
  (\bibinfo{year}{2017}) \bibinfo{pages}{1158--1162}.
\bibitem[{Kyriienko et~al.(2021)Kyriienko, Paine, and Elfving}]{Kyriienko2021}
\bibinfo{author}{O.~Kyriienko}, \bibinfo{author}{A.~E. Paine},
  \bibinfo{author}{V.~E. Elfving},
\newblock \bibinfo{title}{{Solving nonlinear differential equations with
  differentiable quantum circuits}},
\newblock \bibinfo{journal}{Phys. Rev. A} \bibinfo{volume}{103}
  (\bibinfo{year}{2021}) \bibinfo{pages}{052416}.
\bibitem[{Sarma et~al.(2024)Sarma, Watts, Moosa, Liu, and McMahon}]{Sarma2024}
\bibinfo{author}{A.~Sarma}, \bibinfo{author}{T.~W. Watts},
  \bibinfo{author}{M.~Moosa}, \bibinfo{author}{Y.~Liu}, \bibinfo{author}{P.~L.
  McMahon},
\newblock \bibinfo{title}{{Quantum variational solving of nonlinear and
  multidimensional partial differential equations}},
\newblock \bibinfo{journal}{Phys. Rev. A} \bibinfo{volume}{109}
  (\bibinfo{year}{2024}) \bibinfo{pages}{062616}.
\bibitem[{git(2025)}]{github}
\bibinfo{title}{Codes available at
  \url{https://github.com/YYgroup/ChebyMeasure}}, \bibinfo{year}{2025}.
\bibitem[{Kolmogorov(1941)}]{Kolmogorov1941}
\bibinfo{author}{A.~N. Kolmogorov},
\newblock \bibinfo{title}{{The local structure of turbulence in incompressible
  viscous fluid for very large Reynolds numbers}},
\newblock \bibinfo{journal}{Dokl. Akad. Nauk SSSR} \bibinfo{volume}{30}
  (\bibinfo{year}{1941}) \bibinfo{pages}{301--305}.
\bibitem[{Rivlin(2020)}]{Rivlin2020}
\bibinfo{author}{T.~J. Rivlin}, \bibinfo{title}{Chebyshev polynomials},
  \bibinfo{publisher}{Courier Dover Publications}, \bibinfo{year}{2020}.
\bibitem[{Trefethen(2019)}]{Trefethen2019}
\bibinfo{author}{L.~N. Trefethen}, \bibinfo{title}{{Approximation theory and
  approximation practice, extended edition}}, \bibinfo{publisher}{SIAM},
  \bibinfo{year}{2019}.
\bibitem[{Abhari et~al.(2024)Abhari, Treinish, Krsulich, Wood, Lishman, Gacon,
  Martiel, Nation, Bishop, Cross, Johnson, and Gambetta}]{qiskit2024}
\bibinfo{author}{A.~J. Abhari}, \bibinfo{author}{M.~Treinish},
  \bibinfo{author}{K.~Krsulich}, \bibinfo{author}{C.~J. Wood},
  \bibinfo{author}{J.~Lishman}, \bibinfo{author}{J.~Gacon},
  \bibinfo{author}{S.~Martiel}, \bibinfo{author}{P.~D. Nation},
  \bibinfo{author}{L.~S. Bishop}, \bibinfo{author}{A.~W. Cross},
  \bibinfo{author}{B.~R. Johnson}, \bibinfo{author}{J.~M. Gambetta},
\newblock \bibinfo{title}{{Quantum computing with {Q}iskit}},
\newblock \bibinfo{journal}{arXiv.2405.08810}  (\bibinfo{year}{2024}).
\bibitem[{Orszag(1969)}]{Orszag1969}
\bibinfo{author}{S.~A. Orszag},
\newblock \bibinfo{title}{{Numerical methods for the simulation of
  turbulence}},
\newblock \bibinfo{journal}{Phys. Fluids} \bibinfo{volume}{12}
  (\bibinfo{year}{1969}) \bibinfo{pages}{II--250}.
\bibitem[{Graham et~al.(2016)Graham, Kanov, Yang, Lee, Malaya, Lalescu, Burns,
  Eyink, Szalay, Moser, and Meneveau}]{Graham2016}
\bibinfo{author}{J.~Graham}, \bibinfo{author}{K.~Kanov},
  \bibinfo{author}{X.~I.~A. Yang}, \bibinfo{author}{M.~K. Lee},
  \bibinfo{author}{N.~Malaya}, \bibinfo{author}{C.~C. Lalescu},
  \bibinfo{author}{R.~Burns}, \bibinfo{author}{G.~Eyink},
  \bibinfo{author}{A.~Szalay}, \bibinfo{author}{R.~D. Moser},
  \bibinfo{author}{C.~Meneveau},
\newblock \bibinfo{title}{{A web services accessible database of turbulent
  channel flow and its use for testing a new integral wall model for LES}},
\newblock \bibinfo{journal}{Journal of Turbulence} \bibinfo{volume}{17}
  (\bibinfo{year}{2016}) \bibinfo{pages}{181--215}.

\end{thebibliography}

\end{document}